\documentclass[11pt]{article}
\usepackage{latexsym,amssymb,amstext,amsmath}
\usepackage{hyperref}
\allowdisplaybreaks 
\def\oneone{\rlap 1\mkern4mu{\rm l}} % unit matrix
\begin{document}
%%%%%%%%%%%%%%%%%%%%%%  
%
%
\begin{titlepage}
\begin{flushright} \small
 Nikhef-2013-003 \\ ITP-UU-13/03\\AEI-2013-048
\end{flushright}
\bigskip

\begin{center}
  {\LARGE\bfseries Deformations of gauged SO(8) supergravity  \\[2mm]
    and supergravity in eleven dimensions}
  \\[10mm]

{{\bf Bernard de Wit}\\
Nikhef, Science Park 105, 1098 XG Amsterdam, The
Netherlands,  {\it and} \\
Institute for Theoretical Physics, Utrecht University, 
Leuvenlaan 4,\\ 3584 CE Utrecht, The Netherlands\\[.2ex]
{\tt B.deWit@uu.nl}}
\vskip 2mm
{{\bf Hermann Nicolai}\\
Max-Planck-Institut f\"ur Gravitationsphysik (Albert-Einstein-Institut),\\
M\"uhlenberg 1, D-14476 Potsdam, Germany\\[.2ex]
{\tt Hermann.Nicolai@aei.mpg.de}}
\vskip 4mm
\end{center}

\vspace{3ex}

\begin{center}
{\bfseries Abstract}
\end{center}
\begin{quotation} \noindent Motivated by the fact that there exists a
  continuous one-parameter family of gauged $\mathrm{SO}(8)$
  supergravities, possible eleven-dimensional origins of this
  phenomenon are explored. Taking the original proof of the
  consistency of the truncation of $11D$ supergravity to
  $\mathrm{SO}(8)$ gauged supergravity as a starting point, a number
  of critical issues is discussed, such as the preferred
  electric-magnetic duality frame in four dimensions and the existence
  of dual magnetic gauge fields and related quantities in eleven
  dimensions. Some of those issues are resolved but others seem to
  point to obstructions in embedding the continuous degeneracy in
  $11D$ supergravity. While the final outcome of these efforts remains
  as yet inconclusive, several new results are obtained. Among those
  is the full non-linear ansatz for the seven-dimensional flux
  expressed in terms of the scalars and pseudoscalars of $4D$
  supergravity, valid for both the $S^7$ and the $T^7$ truncations
  without resorting to tensor-scalar duality.
\end{quotation}

\vfill

%%%%%%%
%flushleft{\today}
%%%%%%
\end{titlepage}

%%%%%%%%%%%%%%%%%%%%%%%%%%%%%%%%%%%%%%%%%%%%%%%%%%%%%%%%%%%%%%%%%%%
%%%%%%%%%%%%%%%%%%%%%%%%%%%%%%%%%%%%%%%%%%%%%%%%%%%%%%%%%%%%%%%%%%%
\section{Introduction}
\label{sec:introduction}
\setcounter{equation}{0}
%%%%%%%%%%%%%%%%%%%%%%%%%%%%%%%%%%%%%%%%%%%%%%%%%%%%%%%%%%%%%%%%%%%
Recently it was discovered that there exists a continuous
one-parameter family of inequivalent gauged $\mathrm{SO}(8)$
supergravities characterized by one angular parameter $\omega$
\cite{Dall'Agata:2012bb}. The new theories were found by using the
embedding tensor approach
\cite{Nicolai:2001sv,deWit:2002vt,deWit:2007mt} to couple an
$\omega$-dependent linear combination of 28 electric and 28 magnetic
gauge fields and elevate their gauge group to $\mathrm{SO}(8)$. As is
well known one can convert these theories by performing an
$\omega$-dependent electric-magnetic duality transformation so that
the gauging becomes purely electric. The theories thus obtained
correspond to a one-dimensional variety of $N=8$ supergravity Lagrangians 
in which the 28 abelian gauge transformations have
been extended to a non-abelian $\mathrm{SO}(8)$ electric gauge group
in the conventional way; the consistency of this gauging can be
directly inferred by making use of the $T$-tensor identities presented
in \cite{deWit:1982ig}, which remain applicable for non-zero
$\omega$. The inequivalence of the new gauged $\mathrm{SO}(8)$
supergravities for different (generic) values of $\omega$ was
confirmed in \cite{Dall'Agata:2012bb} by examining stationary points
of the potential in a $\mathrm{G}_2$-invariant sector of the theory
which showed that the multiplicities of $\mathrm{SO}(7)$-invariant and
$\mathrm{G}_2$-invariant stationary points are different from those
found for the original gauging
\cite{Warner:1982,deWit:1983gs,Gunaydin:1983mi}. The discovery of the
continuous deformations has meanwhile stimulated further work on more
general solutions of gauged $\mathrm{SO}(8)$ supergravities
\cite{Borghese:2012qm,Borghese:2012zs}.

The existence of a continuous family of gauged $\mathrm{SO}(8)$
supergravities is a rather surprising fact and its discovery
demonstrates the power of the embedding tensor method.  In this paper
we first rederive and clarify this result in the context of the
electric duality frame, following as much as possible the original
construction of the $\mathrm{SO}(8)$ gauging \cite{deWit:1982ig}. The
analysis in the electric frame is interesting in its own right. It
enables us to compare the $\mathrm{SO}(7)^\pm$ solutions that were
found in the electric frame for $\omega=0$
\cite{Warner:1982,deWit:1983gs} to the corresponding solutions in the
$\omega$-deformed theory. Besides confirming the consistency of the
gaugings, it provides an independent verification of the phenomenon,
noted in \cite{Dall'Agata:2012bb}, that the independent deformations
cover only part of the full interval $\omega\in (0,2\pi]$.  In the
electric duality frame this is caused by the fact that certain changes
in $\omega$ can be compensated for by performing various field
redefinitions in the Lagrangian, so that different values of $\omega$
will correspond to the same Lagrangians. Ultimately this reduces the
interval of inequivalent deformations to $\omega\in (0,\pi/8]$.  In
establishing this result the diagonal $\mathrm{SU}(8)$ subgroup of
$\mathrm{E}_{7(7)}\times \mathrm{SU}(8)$ plays an important role,
where $\mathrm{E}_{7(7)}$ is the symmetry group of the ungauged theory
\cite{Cremmer:1979up}.

The prime motivation for our work is to explore whether the continuous
deformation has a possible interpretation from the perspective of
$11D$ supergravity \cite{Cremmer:1978km}, or, more precisely, whether
the deformed theories can be consistently embedded into $11D$
supergravity.  The original gauged $\mathrm{SO}(8)$ supergravity has
been proven to correspond to a consistent truncation of $11D$
supergravity associated with $S^7$ \cite{deWit:1986iy,Nicolai:2011cy};
this proof made use of the $\mathrm{SL}(8)$ invariant formulation of
the $4D$ theory with the $\mathrm{SO}(8)$ gauge group embedded into
$\mathrm{SL}(8)$. Therefore we first address the question whether or
not this proof can be extended to the $\omega$-dependent electric
duality frame. The answer turns out to be negative. Therefore the only
option seems to remain within the context of the $\mathrm{SL}(8)$
covariant duality frame and to investigate whether one can
consistently incorporate the magnetic charges in this frame in the
context of the higher-dimensional theory. As we intend to show in this
paper, the $\mathrm{SU}(8)$ covariant reformulation of $11D$
supergravity given in \cite{deWit:1985iy,deWit:1986mz} does indeed
allow for the necessary dual structures. On the other hand, the
assumption that the $\omega$-deformed theories also have a consistent
embedding in $11D$ supergravity, would imply that any solution of
$11D$ supergravity that is known to have a $4D$ counterpart for
$\omega=0$ will belong to one-parameter family of similar solutions of
$11D$ supergravity. In view of the fact that the $\omega$-deformation
commutes with $\mathrm{SO}(8)$ the solutions belonging to such a
family should share the same invariance subgroup of
$\mathrm{SO}(8)$. For instance, a continuous family should exist of $\mathrm{SO}(7)$
invariant solutions associated with the $11D$ solutions of
\cite{Englert:1982vs,deWit:1984va} that have been shown to correspond
to similar solutions of $4D$ $\mathrm{SO}(8)$-gauged supergravity with
$\omega=0$ \cite{deWit:1983vq,deWit:1986iy,Nicolai:2011cy}. It seems
that this is only possible when $11D$ supergravity is somehow extended
such that it will be equipped with the deformation parameter $\omega$
as an extraneous parameter, which would require an extension of the
version of $11D$ supergravity given in \cite{Cremmer:1978km}. The
nature of such an extension is at present not known. We discuss these
issues in the concluding section \ref{sec:Membedding}.

While a complete resolution of the important question concerning the
possible $11D$ relation of the $\omega$-deformed supergravities
remains open for the moment, the consideration of dual vectors in the
$11D$ context leads us to two unexpected and important new results
which generalize the $\mathrm{SU}(8)$ invariant reformulation of $11D$
supergravity given in \cite{deWit:1985iy,deWit:1986mz} on which the
consistency proof of \cite{deWit:1986iy,Nicolai:2011cy} was based.
The first one is the existence of a new `generalized vielbein' that is
related to the 28 dual magnetic vectors in the same way as the
original generalized vielbein was related to the 28 electric
vectors. More specifically, the latter is a soldering form
$e^m{}_{AB}$ associated to the Kaluza-Klein vector fields $B_\mu{}^m$
(contained in the elfbein $E_M{}^A$ of $11D$ supergravity
(cf. \ref{eq:elfbein})), while the new vielbein $e_{mn\,AB}$ is
associated to the components $A_{\mu mn}$ and $A_{mnp}$ of the
three-form potential $A_{MNP}$ of $11D$
supergravity.~\footnote{%%%%%%%%%%%%%%%%%%%%%%%
  A similar extension has already appeared in a previous study
  \cite{Koepsell:2000xg} in the context of $3D$ supergravity and
  $\mathrm{E}_{8(8)}$, where the vectors are dual to scalar fields,
  but where it is not possible to compare the relevant formulae to
  non-trivial compactifications of $11D$ supergravity. }
%%%%%%%%%%%%%%%%%%%%%%%%%%%%%%%%%%%%%%%%%
The combination of the two generalized vielbeine then yields the
formula (\ref{eq:A-mnp}) for the {\em non-linear flux ansatz},
analogous to the non-linear metric ansatz first presented in
\cite{deWit:1984nz,deWit:1986iy}. A formula for the flux had already
been derived in \cite{deWit:1986iy,Nicolai:2011cy}, but that formula
was in terms of the four-form field strength rather than the
three-form potential and appears to be too unwieldy for practical
applications. This is not so with the new and much simpler formula
\eqref{eq:A-mnp} which is directly in terms of the three-form
potential $A_{mnp}$. It is remarkable that the detour via the
$\omega$-deformed gaugings thus yields the answer to a question that
has remained open for almost 30 years!

This paper is organized as follows. Section \ref{sec:so(8)-gaugings}
summarizes a number of characteristic features of $N=8$ supergravity
and of the relevant electric-magnetic duality frames. Subsequently the
$\omega$-deformed $\mathrm{SO}(8)$ gaugings are discussed in the
electric frame and we analyze the inequivalence of supergravities
corresponding to different values of $\omega$. In section
\ref{sec:so(7)-solutions} an analysis is presented of the
$\mathrm{SO}(7)^\pm$ solutions for arbitrary values of $\omega$. The
results are in agreement with those presented in
\cite{Dall'Agata:2012bb}. In the subsequent section
\ref{sec:embedding} the possible embedding of the $\omega$-deformed
theories is considered. The first conclusion is that such an embedding
can only be given in the $\mathrm{SL}(8)$ duality frame, which implies
that a possible embedding should involve dual magnetic gauge fields as
well as related quantities. The search for such dual quantities is
then undertaken in section \ref{sec:dual-quantities}. Although such
quantities can indeed be identified, it still does not enable the
formulation of a consistent embedding scheme of the $\omega$-deformed
$4D$ theories into $11D$ supergravity. On the other hand the newly
found dual gauge fields and generalized vielbeine give substantial new
insights of the embedding of the original $\omega=0$ theory into $11D$
supergravity. In particular a non-linear expression is found for the
tensor field $A_{mnp}$ of $11D$ supergravity in the $S^7$ and $T^7$
truncations. Conclusions and a further outlook are presented in
section \ref{sec:Membedding}. An appendix
\ref{App:derivation-dual-gen-vielbein} presents a number of
definitions and the algebraic details related to the supersymmetry transformation
rule of the dual generalized vielbein.

%%%%%%%%%%%%%%%%%%%%%%%%%%%%%%%%%%%%%%%%%%%%%%%%%%%%%%%%%%%%%%%%%%%
\section{SO(8) gaugings of maximal D=4 supergravity}
\label{sec:so(8)-gaugings}
%%%%%%%%%%%%%%%%%%%%%%%%%%%%%%%%%%%%%%%%%%%%%%%%%%%%%%%%%%%%%%%%%%%
As is well known, four-dimensional Lagrangians with abelian gauge
fields are ambiguous, as different Lagrangians can lead to equivalent
field equations and Bianchi identities. This phenomenon is known as
electric-magnetic duality. Generic electric-magnetic duality
transformations do not constitute an invariance but an
equivalence. These transformations can be effected by performing a real
symplectic rotation of the field strengths $F_{\mu\nu}$ and the dual
fields strengths $G_{\mu\nu}$. The latter are defined such that the
Bianchi identity on the latter equals precisely the field equations of
the vector fields. For $N=8$ supergravity we have 28 vector fields so
that the number of field strengths and dual field-strengths equals 56.
The general analysis of \cite{Gaillard:1981rj} therefore implies that
the electric-magnetic duality group is equal to
$\mathrm{Sp}(56;\mathbb{R})$. After applying the symplectic rotation
of the field strengths, the new dual field strengths $G_{\mu\nu}$ take
a different form that will in turn follow from a different
Lagrangian. In the absence of a gauging, all these Lagrangians are
physically equivalent as they describe the same set of field equations
and Bianchi identities.

The corresponding theory may in principle be invariant under a
subgroup of the electric-magnetic dualities combined with related
transformations on the other fields, meaning that the Lagrangian will
not change under this subgroup (which does not imply that the
Lagrangian is invariant in the naive sense, as the Lagrangian does not
transform as a function under duality). This happens for ungauged
$N=8$ supergravity where the invariance group corresponds to the
non-compact $\mathrm{E}_{7(7)}$ subgroup of
$\mathrm{Sp}(56;\mathbb{R})$ \cite{Cremmer:1979up}. When working with
a formulation that is gauge invariant under local chiral
$\mathrm{SU}(8)$, which acts on the fermions and on the scalars, the
theory is invariant under the group $\mathrm{E}_{7(7)}\times
\mathrm{SU}(8)$ which is linearly realized. Once a gauge is adopted
with respect to the local $\mathrm{SU}(8)$, the group action of
$\mathrm{E}_{7(7)}$ will be non-linearly realized on the spinors and
the scalars of the theory. The latter then parametrize an
$\mathrm{E}_{7(7)}/\mathrm{SU}(8)$ coset space; here it is relevant
that $\mathrm{SU}(8)$ is the maximal compact subgroup of
$\mathrm{E}_{7(7)}$. We prefer to work with the linear version of the
theory with manifest local $\mathrm{SU}(8)$ invariance.

However, the Lagrangian can only be invariant under a subgroup of
$\mathrm{E}_{7(7)}$, such as, for instance, $\mathrm{SL}(8)$, under
which the vector fields transform in the real $\boldsymbol{28}$
representation. While the usefulness of real representations is
obvious for the gauge fields, it is not convenient for the remaining
fields which transform under $\mathrm{SU}(8)$ in complex
representations. A crucial quantity in the formulation of the theory
is the so-called 56-bein $\mathcal{V}$, which is a $56\times56$ matrix
that belongs to the $\boldsymbol{56}$ representation of
$\mathrm{E}_{7(7)}$. The usual representation of this matrix is given
in a pseudo-real decomposition of $\mathrm{E}_{7(7)}$ based on
$\boldsymbol{56}= \boldsymbol{28}+\overline{\boldsymbol{28}}$, where
$\boldsymbol{28}$ and $\overline{\boldsymbol{28}}$ denote two
conjugate representations of the maximal subgroup
$\mathrm{SU}(8)$. The 56-bein $\mathcal{V}$ will transform under
$\mathrm{E}_{7(7)}$ rigid transformations and under lcoal
$\mathrm{SU}(8)$ by right- and left-multiplication,
respectively.\footnote{%%%%%%%%%%%%%%%%%%%%%%%%%%%%%%%%%%%%%%%%
  There are different conventions used in the literature. Here we will
  follow \cite{deWit:1982ig}.} %%%%%%%%%%%%%%%%%%%%%%%%%%%%%%%%%%%

To set the stage let us briefly discuss some properties of the group
$\mathrm{E}_{7(7)}\subset \mathrm{Sp}(56;\mathbb{R})$. We start with
the fundamental representation $\boldsymbol{56}$ of
$\mathrm{Sp}(56;\mathbb{R})$, written as a pseudo-real vector
$(z_{IJ}, z^{KL})$ with $z^{IJ} = (z_{IJ})^\ast$, where the indices
are anti-symmetric index pairs $[IJ]$ and $[KL]$ and $I,J,K,L =
1,\ldots,8$. Hence the $(z_{IJ},z^{KL})$ span a real 56-dimensional
vector space. Consider infinitesimal transformations of the form,
\begin{align}
  \label{eq:delta-z}
\delta z_{IJ} =&\, \Lambda_{IJ}{}^{\!KL} \,z_{KL} + \Sigma_{IJKL}
\,z^{KL}\,,\nonumber\\
\delta z^{IJ} =&\, \Lambda^{IJ}{}_{\!KL} \,z^{KL} + \Sigma^{IJKL}
\,z_{KL}\,.
\end{align}
where $\Lambda_{IJ}{}^{\!KL}=\Lambda_{[IJ]}{}^{\![KL]}$ and
$\Sigma_{IJKL}= \Sigma_{[IJ]\,[KL]}$ are subject to the conditions,
\begin{equation}
  \label{eq:sp56-properties}
(\Lambda_{IJ}{}^{\!KL})^\ast =  \Lambda^{IJ}{}_{\!KL}=
-\Lambda_{KL}{}^{\!IJ} \,, \qquad
   (\Sigma_{IJKL})^\ast = \Sigma^{IJKL}= \Sigma^{KLIJ} \,.
\end{equation}
Note that complex conjugation is effected by raising or lowering of
indices.  The corresponding group elements $g$ constitute the group
${\rm Sp}(56;\mathbb{R})$ in a pseudo-real basis provided that they
satisfy the conditions,
\begin{equation}
  \label{eq:def-sp56}
  g^\ast = \omega \,g\, \omega\,, \qquad g^{-1} = \Omega\, g^\dagger
  \,\Omega\,,   
\end{equation}
where $\omega$ and $\Omega$ are given by 
\begin{equation}
  \label{eq:omega-matrices}
  \omega= \begin{pmatrix}0& \oneone \\ \oneone & 0 \end{pmatrix} \;,\qquad 
\Omega= \begin{pmatrix} \oneone &0 \cr 0& -\oneone \end{pmatrix}  \;.  
\end{equation}
The above properties ensure that the sesquilinear form, $(z_1,z_2) =
z_1^{IJ}\,z_{2IJ} - z_{1IJ}\,z_2^{IJ}$, is invariant. The generators
associated with $\Lambda_{IJ}{}^{KL}$ generate the maximal compact
$\mathrm{U}(28)$ subgroup of ${\rm Sp}(56;\mathbb{R})$, and a 
$\mathrm{GL}(28)$ subgroup is generated by real matrices
$\Lambda_{IJ}{}^{KL}$ and purely real or purely imaginary
$\Sigma^{IJKL}$, whose compact subgroup equals $\mathrm{SO}(28)$.

Let us now consider the ${\rm E}_{7(7)}$ subgroup, for which 
$\Sigma^{IJKL}$ is fully anti-symmetric and the generators are further
restricted according to  
\begin{align}
  \label{eq:E7}
& \Lambda_{IJ}{}^{\!KL} = \delta_{[I}^{[K} \,\Lambda_{J]}{}^{\!L]}  \,, 
\qquad \Lambda_I{}^J =  -\Lambda^J{}_I\,, \nonumber\\
&
\Lambda_I{}^I = 0\,,\qquad \Sigma_{IJKL} = \tfrac1{24}
\varepsilon_{IJKLMNPQ} \,\Sigma^{MNPQ}\,.
\end{align}
Obviously the matrices $\Lambda_{I}{}^{\!J}$ generate the group ${\rm
  SU}(8)$, which has dimension 63; since the $\Sigma_{IJKL}$ comprise
70 real parameters, the dimension of ${\rm E}_{7(7)}$ equals
$63+70=133$. Because ${\rm SU}(8)$ is the maximal compact subgroup,
the number of non-compact generators minus the number of compact ones
equals $70-63=7$. It is straightforward to show that these matrices
close under commutation and generate the group ${\rm E}_{7(7)}$
\cite{Cremmer:1979up,deWit:1982ig}. To show this one needs a variety
of identities for self-dual tensors. Note that ${\rm E}_{7(7)}$ has
another maximal 63-dimensional subgroup, which is real but not
compact, namely the group $\mathrm{SL}(8)$. It is generated by those
matrices in (\ref{eq:E7}) for which the sub-matrices
$\Lambda_I{}^{\!J}$ and $\Sigma^{IJKL}$ are both real.

Let us now define the 56-bein $\mathcal{V}$, which describes the
scalar fields,  
\begin{equation}
  \label{eq:u-v}
  \mathcal{V}(x) =
  \begin{pmatrix} u_{ij}{}^{\!IJ}(x) &  v_{ ij\,KL}(x)\\[6mm]
    v^{kl\,IJ}(x) &   u^{kl}{}_{\!KL}(x) \end{pmatrix} \,,
\end{equation}
and which is an element of $\mathrm{E}_{7(7)}$. Therefore it can
transform by left-multiplication under local $\mathrm{SU}(8)$ and by
right-multiplication under rigid $\mathrm{E}_{7(7)}$. Hence the
indices $[ij]$ and $[kl]$ are local $\mathrm{SU}(8)$ indices and
$[IJ]$ and $[KL]$ are rigid $\mathrm{E}_{7(7)}$ indices. A standard
$\mathrm{SU}(8)$ gauge condition leads to the following coset
representative (`unitary gauge'), 
\begin{equation}
  \label{eq:u-v-phi}
  \mathcal{V}(x) = \exp 
  \begin{pmatrix} 0 &  -\tfrac14\sqrt{2} \,\phi_{ ijkl}(x)\\[6mm]
     -\tfrac14\sqrt{2}\,\phi^{mnpq}(x) &   0  \end{pmatrix} \,,
\end{equation}
where the $\phi^{ijkl}$ are complex fields transforming as an anti-symmetric
four rank tensor under the linearly realized rigid
$\mathrm{SU}(8)$. The complex conjugate fields, $\phi_{ijkl}$, 
are related to the original fields by a complex self-duality constraint,
\begin{equation}
  \label{eq:selfdual-phi}
  \phi_{ijkl}= \tfrac1{24}\varepsilon_{ijklmnpq}\,\phi^{mnpq}\,.   
\end{equation}
Observe that in this gauge the indices $I,J,K,\ldots$ are no longer
distinguishable from the $\mathrm{SU}(8)$ indices $i,j,k,\ldots$. 
We also note that the reflection $\phi^{ijkl} \to - \phi^{ijkl}$ maps
$(u,v) \to (u,-v)$ in (\ref{eq:u-v}) and therefore corresponds to a 
trivial reparametrization of the E$_{7(7)}/$SU(8) coset space.

Subsequently we consider the 28 field strengths $F_{\mu\nu}{}^{IJ}$
and their dual field strengths,
\begin{equation}
  \label{eq:G}
  G^{+\mu\nu}{}_{IJ} = -\frac4{e}\, \frac{\partial
    \mathcal{L}}{\partial F^+{}_{\mu\nu IJ}} \,. 
\end{equation}
The Bianchi identies and the field equations of the vector fields are
summarized in the following equations,
\begin{equation}
  \label{eq:Bianchi-field-eq}
  \partial_\mu\big[e\,F^{+\mu\nu IJ} - e \, F^{-\mu\nu IJ}\big]
  =0 =
  \partial_\mu\big[e\, G^{+\mu\nu}_{IJ} + e \, G^{-\mu\nu}{}_{IJ}\big]\,.  
\end{equation}
These equations can be written in terms of a 56-component array of
selfdual field strengths, $(F^+_{1\mu\nu IJ}, F^+_{2\mu\nu}{}^{IJ})$,
defined by
\begin{align}
  \label{eq:F-1-2}
   F^+_{1\mu\nu}{}_{IJ} = &\, \tfrac12\big(G^{+\mu\nu}{}_{IJ}+
   F^{+\mu\nu}{}^{IJ} \big)\,,\nonumber\\ 
   F^+_{2\mu\nu}{}^{IJ} =&\, \tfrac12\big(G^{+\mu\nu}{}_{IJ}-
   F^{+\mu\nu{IJ}} \big)\,, 
\end{align}
and their anti-selfdual ones $(F^-_{1\mu\nu}{}^{IJ},
F^-_{2\mu\nu IJ})$ that follow by complex conjugation, in a form that
is manifestly covariant under $\mathrm{Sp}(56;\mathbb{R})$
\cite{Gaillard:1981rj}.

What remains is to specify $G_{\mu\nu{IJ}}$ in terms of
$F_{\mu\nu}{}^{IJ}$ and terms depending on the matter fields. This
will then determine all terms involving the vector fields of the
Lagrangian. As long as we have not switched on the gauging, the matter
field contributions come exclusively from fermionic bilinears, which
we denote by $\mathcal{O}_{\mu\nu}$. Since the fermions transform under local
$\mathrm{SU}(8)$ and not under $\mathrm{E}_{7(7)}$, this relation must
necessarily involve the 56-bein $\mathcal{V}$ and can be written as follows
\cite{deWit:1982ig}, 
\begin{equation}
  \label{eq:FGO}
  \mathcal{V}\, \begin{pmatrix} F_{1\mu\nu\,IJ}^{+}
    \\[2mm]  F^+_{2\mu\nu}{}^{KL} \end{pmatrix}   =
  \begin{pmatrix} \bar F_{\mu\nu\,ij}^{+}\\[3mm]
    {\cal O}_{\mu\nu}^{+kl} \end{pmatrix} \,,
\end{equation}
where ${\cal O}^+_{\mu\nu}{}^{ij}$ is an ${\rm SU}(8)$ covariant
tensor quadratic in the fermion fields and independent of the scalar
fields, which appears as a moment coupling in the Lagrangian. Without
going into the details we mention that chirality and self-duality
restricts the form of $\mathcal{O}^+_{\mu\nu}{}^{ij}$ up to some
normalization constants.  The tensor $\bar F^+_{\mu\nu\,ij}$ is an
${\rm SU}(8)$ covariant field strength which appears in the
supersymmetry transformation rules of the spinors, which is simply
defined by the above condition. For future reference we give the
definition of $\mathcal{O}^+_{\mu\nu}{}^{ij}$,
\begin{align}
  \label{eq:def-O}
  \mathcal{O}^+_{\mu\nu}{}^{ij}=&\,- \tfrac1{288}\sqrt{2}\,
  \varepsilon^{ijklmnpq} 
  \,\bar \chi_{klm}\gamma_{\mu\nu}\chi_{npq}  - \tfrac1{4}\bar\psi_{\rho k}
  \gamma_{\mu\nu} \gamma^\rho \chi^{ijk} + \tfrac1{4}\sqrt{2}\, 
  \bar\psi_\rho{}^i \gamma^{[\rho}\gamma_{\mu\nu}\gamma^{\sigma]}
  \psi_\sigma{}^j\,.
\end{align}

The form of \eqref{eq:FGO} emphasizes the covariance under the group
$\mathrm{SL}(8)$, as both $F^+_{1\mu\nu IJ}$ and
$F^+_{2\mu\nu}{}^{IJ}$ defined in \eqref{eq:F-1-2} transform in the
$\boldsymbol{28}$ and $\overline{\boldsymbol{28}}$ representations of
that group. As long as we have not switched on the gauging, we have
the option of changing the basis of these field strengths by a matrix
$E\in\mathrm{Sp}(56;\mathbb{R})$. It thus seems that the possible
Lagrangians are encoded in these matrices $E$. However, this is not
the case, because, when $E$ belongs to $\mathrm{GL}(28)$ or to
$\mathrm{E}_{7(7)}$, it can be absorbed into either the field
strengths \eqref{eq:F-1-2} or into the 56-bein, respectively. Hence it
follows that \eqref{eq:FGO}, and thus the Lagrangian has an ambiguity
encoded in a matrix \cite{deWit:2002vt,deWit:2002vz}
\begin{equation}
E \; \in \; {\rm E}_{7(7)} \backslash {\rm
  Sp}(56;\mathbb{R})/\mathrm{GL}(28,\mathbb{R})\,.
\end{equation}  
When one is interested in $\mathrm{SO}(8)$ invariant Lagrangians, the
matrix $E$ must preserve the $\mathrm{SO}(8)$ subgroup, so that the
relevant matrices $E$ are restricted to
\begin{equation}
  \label{eq:E-angle}
  E = \begin{pmatrix} \mathrm{e}^{\mathrm{i} \omega} \oneone & 0\\[3mm]
    0& \mathrm{e}^{-\mathrm{i} \omega}\oneone  \end{pmatrix}\, ,
\end{equation}
where $\oneone \equiv \oneone_{28}$ denotes the $28\times28$ unit
matrix. Hence these Lagrangians are encoded in a single angle
$\omega$. \footnote{%%%%%%%%%%%%%%%
  Angles such as $\omega$ were first introduced in the context of
  gauged $N=4$ supergravity in \cite{deRoo:1985jh}} %%%%%%%%%%%%%
For special values of $\omega$ this matrix will constitute an element
of $\mathrm{E}_{7(7)}$, because the compact $\mathrm{SU}(8)$ subgroup
of $\mathrm{E}_{7(7)}$ has a non-trivial center $Z[\mathrm{SU}(8)] =
\mathbb{Z}_8$, which is reduced to $\mathbb{Z}_4$ when acting on
bosons (as these come with an even number of SU(8) indices). The
center $Z[\mathrm{SU}(8)]$ consists of the matrices
$\mathrm{e}^{\mathrm{i}\omega/2} \oneone_{8}$ with $\omega$ a multiple
of $\pi/2$. Consequently, $\mathrm{SO}(8)$ invariant Lagrangians
corresponding to $\omega$-values that differ by an integer times
$\pi/2$ must be equivalent, as they are related by an element of
$\mathrm{SU}(8)$ (and therefore of $\mathrm{E}_{7(7)}$). Other than
these there are no matrices $E$ belonging to $\mathrm{E}_{7(7)}$. We
return momentarily to a more detailed analysis of possible
equivalences.

The exponential factor in \eqref{eq:E-angle} can now be incorporated
directly into the supergravity Lagrangian by simply including
$\omega$-dependent phase factors into the submatrices $u$ and $v$ in
the Lagrangian according to
\begin{equation}
  \label{eq:uv-redef}
  u_{ij}{}^{IJ}\to \mathrm{e}^{\mathrm{i} \omega}
  u_{ij}{}^{IJ}\,,\quad  v_{ijIJ}\to \mathrm{e}^{-\mathrm{i} \omega}
  v_{ijIJ}\,. 
\end{equation}
This defines the deformed supergravity Lagrangians in the electric
frame. As already mentioned in section~\ref{sec:introduction}, the
inequivalent theories do not cover the full interval $\omega\in
(0,2\pi]$, but are restricted to the smaller interval
$\omega\in(0,\pi/8]$, as was shown by \cite{Dall'Agata:2012bb} in a
mixed electric-magnetic duality frame. We will now verify this result
in the electric frame. We distinguish three types of equivalence
transformations for $\omega$: %%%%%%%%%%%%
\begin{enumerate}
\item[i)] The shift $\omega\to\omega+\pi/2$, which can be undone by a
  special $\mathrm{SU}(8)$ transformation belonging to
  $Z[\mathrm{SU}(8)]$.
\item[ii)] The shift $\omega\to\omega+\pi/4$, which can be undone by
  an $\mathrm{SU}(8)$ transformation that belongs to a square root of
  an element of $Z[\mathrm{SU}(8)]$ accompanied by a linear
  redefinition of the gauge fields $A_\mu{}^{IJ}$.
\item[iii)] The reflection $\omega\to -\omega$, which can be undone by
  a parity transformation.
\end{enumerate}
%%%%%%%%%%%%%%%%%%%%%%%%%%%%%%%%%
To analyze these three equivalences we consider the $\omega$-deformed
Lagrangians. The terms that involve the field strengths are encoded in
\eqref{eq:FGO} subject to the deformation \eqref{eq:uv-redef}. Writing
this equation in terms of the separate components, one obtains
\begin{align}
  \label{eq:FGO-comp} %{i-suppressed}
  \big(u^{ij}{}_{IJ} +\mathrm{e}^{2\mathrm{i}\omega} v^{ijIJ}\big)
  G^+_{\mu\nu{IJ}}=&\, \big(u^{ij}{}_{IJ}
  -\mathrm{e}^{2\mathrm{i}\omega} v^{ijIJ}\big) F^+_{\mu\nu}{}^{IJ} +
  2\,\mathrm{e}^{\mathrm{i}\omega}\mathcal{O}^+_{\mu\nu}{}^{ij} \,,
  \nonumber\\
  2\, \mathrm{e}^{-\mathrm{i}\omega} \bar F^+_{\mu\nu{ij}} =&\,
  \big(u_{ij}{}^{IJ} +\mathrm{e}^{-2\mathrm{i}\omega} v_{ijIJ}\big)
  G^+_{\mu\nu{IJ}} +\big(u_{ij}{}^{IJ} -\mathrm{e}^{-2\mathrm{i}\omega}
  v_{ijIJ}\big) F^+_{\mu\nu}{}^{IJ} \,. 
\end{align}

Let us first consider the effect of the shift $\omega\to\omega+\pi/2$
in \eqref{eq:FGO-comp}, which we can clearly undo by performing the
following redefinitions,
\begin{equation}
  \label{eq:redef-v-O-Fbar}
  v^{ijIJ}\to \mathrm{e}^{-\mathrm{i}\pi} v^{ijIJ} = - v^{ij IJ} \,, \quad
  \mathcal{O}^+_{\mu\nu}{}^{ij} 
  \to \mathrm{e}^{-\mathrm{i}\pi/2} \mathcal{O}^+_{\mu\nu}{}^{ij}\,,
  \quad \bar F^+_{\mu\nu{ij}} \to \mathrm{e}^{\mathrm{i}\pi/2} \bar
  F^+_{\mu\nu{ij}} \,. 
\end{equation}
We have to ensure that these redefinitions are consistent for the full
Lagrangian. This follows rather straightforwardly by noting that the
redefinitions \eqref{eq:redef-v-O-Fbar} are precisely generated by
applying a uniform $\mathrm{SU}(8)$ transformation belonging to the
diagonal subgroup of $\mathrm{E}_{7(7)}\times \mathrm{SU}(8)$ and
equal to $\mathrm{e}^{\mathrm{i}\pi/2}\,\oneone_{28}$, which
constitutes an element of $Z[\mathrm{SU}(8)]$. Note that on
$u^{ij}{}_{IJ}$ the effect of this transformation cancels, as it acts
on both index pairs $[ij]$ and $[IJ]$, while it correctly accounts for
the phase factor in the redefinition of $v^{ijIJ}$.~\footnote{ %%%%%%%
  Note that the diagonal $\mathrm{SU}(8)$ transformations induce a
  corresponding change on the field $\phi^{ijkl}$ in the coset
  representative \eqref{eq:u-v-phi}. For this reason the
  pseudo-reality constraint \eqref{eq:selfdual-phi} will be preserved
  throughout. } %%%%%%%%%%%%%%%%%%%%%%%
The $\mathrm{SU}(8)$ transformation is also realized on the fermions
where it takes the form,
\begin{equation}
  \label{eq:fermion-redef}
  \psi_\mu{}^i \to \mathrm{e}^{-\mathrm{i}\pi/4} \psi_\mu{}^i
  \,,\qquad \chi^{ijk} \to  \mathrm{e}^{-3\mathrm{i}\pi/4} \chi^{ijk}
  \,,
\end{equation}
and this generates the desired redefinition of
$\mathcal{O}^+_{\mu\nu}{}^{ij}$ and $\bar F^+_{\mu\nu ij}$.  As far as
the ungauged Lagrangian and the supersymmetry transformations are
concerned (we remind the reader that $\bar F^+_{\mu\nu{ij}}$ and its
anti-selfdual component appear in the supersymmetry transformations),
the shift $\omega\to\omega+\pi/2$ combined with a special
$\mathrm{SU}(8)$ transformation leaves the Lagrangian and the
supersymmetry transformations unaffected. Note that the fact that
the Lagrangian and the supersymmetry transformations are consistent
with respect to {\it local} $\mathrm{SU}(8)$ plays a crucial role for
the remaining terms in the Lagrangian.

To prove that the terms depending on the $\mathrm{SO}(8)$ gauging are
not affected by the shift and the various field redefinitions, we
consider the so-called $T$-tensor associated with the $\mathrm{SO}(8)$
gauging, which takes the following form in the $\omega$-deformed
theory,
\begin{align}
  \label{eq:def-T-tensor}
  T_i{}^{jkl} (\omega\,;\,u,v) =&\,
  \big(\mathrm{e}^{-\mathrm{i}\omega} \,u^{kl}{}_{IJ}  
   +\mathrm{e}^{\mathrm{i} \omega} \,v^{kl IJ}\big)
   \,\big(u_{im}{}^{JK}\, u^{jm}{}_{KI} -v_{imJI}\, v^{jmKL}  \big)
   \nonumber\\[1mm]
   =&\, \cos \omega \, T^{\rm (e)}{}_{\! i}{}^{jkl}(u,v) \, + \, 
   \sin \omega \, T^{\rm (m)}{}_{\! i}{}^{jkl} (u,v)    \,. 
\end{align} 
where in the second line, we explicitly display the decomposition of
the $T$-tensor into an `electric' and a `magnetic' component. As the
reader can check, the consistency of the gauging is not affected by
the $\omega$-deformation \eqref{eq:uv-redef}, because the analysis given in
\cite{deWit:1982ig} still applies, in the sense that all the
`$T$-identities' remain valid.~\footnote{These identities encode the
  same information as the linear and quadratic identities that the
  embedding tensor has to satisfy.}  This is consistent with the
general outline given in \cite{deWit:2002vt,deWit:2007mt} and the
specific application described in \cite{Dall'Agata:2012bb}. When
applying the shift $\omega\to \omega+\pi/2$ in \eqref{eq:def-T-tensor}
we follow the same strategy as before and obtain the relation,
\begin{equation}
  \label{eq:phase-T}
  T_i{}^{jkl} \left(\omega+ \pi/2 \,;\, u,v\right) \; = \;
  \mathrm{e}^{-\mathrm{i}\pi/2} \, T_i{}^{jkl}(\omega\,;\, u,\mathrm{e}^{\mathrm{i}\pi}\,v)\,, 
\end{equation}
where $u$ and $v$ denote $u^{ij}{}_{IJ}$ and $v^{ijIJ}$, respectively.
Again the changes take the form of an $\mathrm{SU}(8)$ transformation,
and are precisely cancelled by the redefinitions found previously in
\eqref{eq:redef-v-O-Fbar} and \eqref{eq:fermion-redef}.

The discussion of the second equivalence transformation $\omega\to
\omega+\pi/4$ proceeds along the same lines, but there are new
features. First of all, because the transformation
$\mathrm{e}^{\mathrm{i}\pi/8}\,\oneone_{8}$ is clearly not an element
of $\mathrm{SU}(8)$, we must replace the identity matrix in this
product by some other real matrix $P_{8}$. Hence we consider
$\mathrm{e}^{\mathrm{i}\pi/8}\,P_{8}$, which constitutes an element of
$\mathrm{SU}(8)$ provided that $P_8$ is real and orthogonal with
$\det[P_8]=-1$. As its square should belong to $Z[\mathrm{SU}(8)]$, it
follows also that $(P_8)^2= \oneone_{8}$. Obviously such matrices
$P_8$ exist!  Examples are diagonal matrices with $p$ eigenvalues
equal to $-1$ and $8-p$ eigenvalues equal to $+1$, with $p$ odd, but
there exist more matrices that satisfy these requirements. The
$\mathrm{SU}(8)$ transformation can also be written in the
$\boldsymbol{28}$ representation, where it takes the form
$\mathrm{e}^\mathrm{\mathrm{i}\pi/4} \, \Pi$, with $\Pi^{ij}{}_{kl} =
P_8{}^{[i}{\!}_{[k}\,P_8{}^{j]}{}_{l]}$.

Now let us return to \eqref{eq:FGO-comp}, but now multiplied by the
matrix $\Pi$ from the left. Furthermore we multiply the field
strength tensors with $\Pi^2=\oneone_{28}$. Obviously the shift in
$\omega$ can now be absorbed by making the following redefinitions,
\begin{equation}
  \label{eq:redef-v-O-Fbar-sqrt}
  \begin{array}{rcl} 
  u^{ij}{}_{IJ} &\!\!\to&\!\! \Pi^{ij}{}_{kl} \,u^{kl}{}_{KL} \,
  \Pi^{KL}{}_{IJ}\,, \\[.2ex]
  v^{ijIJ} &\!\!\to&\!\! \mathrm{e}^{-\mathrm{i}\pi/2}\,\Pi^{ij}{}_{kl}\,
  v^{klKL} \,\Pi^{IJ}{}_{KL}  \,, 
  \end{array} 
  \qquad 
  \begin{array}{rcl} 
  \mathcal{O}^+_{\mu\nu}{}^{ij}   &\!\!\to&\!\!
  \mathrm{e}^{-\mathrm{i}\pi/4} \, \Pi^{ij}{}_{kl} \,\mathcal{O}^+_{\mu\nu}{}^{kl}\,, \\[.2ex]
  \bar F^+_{\mu\nu{ij}} &\!\!\to&\!\! \mathrm{e}^{\mathrm{i}\pi/4}
  \,\Pi^{kl}{}_{ij} \, \bar  F^+_{\mu\nu{kl}} \,. 
  \end{array}
\end{equation}
combined with a linear redefinition of the vector gauge fields, 
\begin{equation}
  \label{eq:redef-A}
  A_\mu{}^{IJ} \to  \Pi^{IJ}{}_{KL}\, A_\mu{}^{KL}\,. 
\end{equation}
The latter induces the same redefinition of the field strengths
$G^+_{\mu\nu IJ}$ and $F^+_{\mu\nu}{}^{IJ}$, even in the presence of
the non-abelian completion. Obviously the transformations
\eqref{eq:redef-v-O-Fbar-sqrt} correspond to $\mathrm{SU}(8)$
transformations belonging to the diagonal subgroup of
$\mathrm{SU}(8)\times \mathrm{E}_{7(7)}$, just as before. On
the fermions they act according to
\begin{equation}
  \label{eq:fermion-redef-sqrt}
  \psi_\mu{}^i \to \mathrm{e}^{-\mathrm{i}\pi/8} \,P_8{}^i{}_j\,\psi_\mu{}^j
  \,,\qquad \chi^{ijk} \to
  \mathrm{e}^{-3\mathrm{i}\pi/8}\,P_8{}^i{}_l \,P_8{}^j{}_m \,P_8{}^k{}_n\,
  \chi^{lmn}   \,.
\end{equation}
For completeness we consider also the change of the $T$-tensor under
the $\omega\to \omega+\pi/4$ transformation, 
\begin{equation}
  \label{eq:phase-T-eq-pi4}
  T_i{}^{jkl} \left(\omega+ \pi/4 \,;\, u,v\right) \; = \;
  \mathrm{e}^{-\mathrm{i}\pi/4} \, P_8{}^m{}_i \, P_8{}^j{}_n \,
  P_8{}^k{}_p \, P_8{}^l{}_q \,  T_m{}^{npq}(\omega\,;\, \Pi \,u\,\Pi,
  \mathrm{e}^{\mathrm{i}\pi/2} \,\Pi \,v\,\Pi)\,,
\end{equation}
with $u$ and $v$ as defined below \eqref{eq:phase-T}.  As a result the
redefinitions noted above cancel precisely the effect of the shift in
$\omega$, which establishes the equivalence in the same fashion as
before.

Finally we consider the third equivalence relation, $\omega\to
-\omega$, whose effect can be absorbed by performing parity reversal
on the fields. To explain this we note that original gauged
$\mathrm{SO}(8)$ supergravity is invariant under parity. Under this
discrete symmetry anti-selfdual and selfdual field strengths are
interchanged simultaneously with the exchange of positive- and
negative-chiral fermion components and of scalar fields with their
complex conjugates. The $\omega$-deformation breaks the invariance
under parity. More precisely, when applying parity reversal to the
Lagrangian for finite $\omega$ one obtains the same Lagrangian with
$\omega$ replaced by $-\omega$. Hence, theories related by $\omega\to
-\omega$ are equivalent, as the sign change can be undone by applying
a parity transformation directly on the fields. Note that the sign
change will also apply to the $T$-tensor given in
\eqref{eq:def-T-tensor}, showing that the magnetic embedding tensor
will change sign.

The three equivalence transformations analyzed in this section imply
that inequivalent Lagrangians are encoded by values of $\omega$ in the
restricted interval $\omega\in(0, \pi/8]$. This result, derived in the
electric frame, is in full agreement with \cite{Dall'Agata:2012bb},
where a fixed duality frame is used and where $\omega$ encodes the
mixture of the electric and magnetic components of the embedding
tensor. In the next section we will analyze the solutions that are
invariant under an $\mathrm{SO}(7)^\pm$ subgroup of the
$\mathrm{SO}(8)$ gauge group. As we shall demonstrate those solutions
reflect precisely the equivalences exhibited in this section.

%%%%%%%%%%%%%%%%%%%%%%%%%%%%%%%%%%%%%%%%%%%%%%%%%%%%%%%%%%%%%%%%%%%%
\section{The potential and  $\mathrm{SO}(7)^\pm$ invariant solutions}
\label{sec:so(7)-solutions}
\setcounter{equation}{0}
%%%%%%%%%%%%%%%%%%%%%%%%%%%%%%%%%%%%%%%%%%%%%%%%%%%%%%%%%%%%%%%%%%%%
The potential of the gauged theory is constructed from the $T$-tensor. 
We recall that this tensor can generally be decomposed into two irreducible
$\mathrm{SU}(8)$ tensors,
\begin{equation}
  \label{eq:decom-T}
  T_i{}^{jkl}= -\tfrac32 A_1{}^{j[k} \,\delta^{l]}{}_i - \tfrac34
  A_{2\,i}{}^{jkl}  \,,
\end{equation}
where $A_1{}^{ij}$ is symmetric in $(ij)$ and $A_{2\,i}{}^{jkl}$ is
anti-symmetric in $[jkl]$ and traceless, $A_{2\,i}{}^{ikl}=0$;
together, these two irreducible components can be assigned to the
$\bf{912}$ of $\mathrm{E}_{7(7)}$ \cite{deWit:1983gs}.  The scalar potential
equals
\begin{equation}
  \label{eq:potential}
  \mathcal{P}=  g^2\left[- \tfrac34 \vert A_1{}^{ij}\vert^2
  +\tfrac1{24} \vert A_{2\,i}{}^{jkl}\vert^2 \right]\,,
\end{equation}
where $g$ is the $\mathrm{SO}(8)$ gauge coupling constant. As shown in
\cite{deWit:1983gs}, this potential has a stationary point whenever
$4\, A_{1\,m[i}\,A_2{}^m{}_{jkl]} -
3\,A_2{}^m{}_{n[ij]}\,A_2{}^n{}_{kl]m}$ is an {\em anti}-selfdual
tensor. 

The simplest examples of special scalar field configurations for which
stationary points exist, and where the effect of the
$\omega$-deformation can be studied in detail, are the backgrounds
preserving $\mathrm{SO}(7)^\pm$-invariance
\cite{Warner:1982,deWit:1983gs,Gunaydin:1983mi}. For these the
56-bein takes the form
\begin{equation}
  \label{eq:u-v-t}
  \mathcal{V}(t) = \exp
  \begin{pmatrix} 0 &  \alpha t\, C^\pm_{ ij\,KL}\\[6mm]
    \alpha^* t \, C^{\pm{kl\,IJ}} &   0  \end{pmatrix} \,,
\end{equation}
with $t\in\mathbb{R}$ and $\alpha = 1$ for $\mathrm{SO}(7)^+$, and
$\alpha=\mathrm{i}$ for $\mathrm{SO}(7)^-$. Here the
$\mathrm{SO}(7)^\pm$ invariant tensors are (anti-)selfdual,
\begin{equation}
  \label{eq:tilde-C}
  C_{IJKL}^\pm = \pm \frac1{24} \varepsilon_{IJKLMNPQ}\, C^\pm_{MNPQ}\,, 
\end{equation}
and obey the condition, 
\begin{equation}
  \label{eq:CC}
  C^\pm_{IJMN} \,C^\pm_{MNKL} = 12\, \delta_{IJ}{}^{KL} \pm 4\, C^\pm_{IJKL}\,.
\end{equation}
Note that \eqref{eq:u-v-t} denotes the coset representative so that we
make no distinction between rigid $\mathrm{SL}(8)$ indices
$I,J,\ldots$ and local $\mathrm{SU}(8)$ indices $i,j,\ldots$. Note
also that field $\phi^{ijkl}$ appearing in \eqref{eq:u-v-phi} is just
equal to $-2\sqrt{2}\,t\,C^{+ijkl}$ or $2\sqrt{2}\,\mathrm{i}
t\,C^{-ijkl}$, respectively, so that the pseudo-reality relation
\eqref{eq:selfdual-phi} is satisfied.

Using the relations \eqref{eq:tilde-C} and \eqref{eq:CC}, one shows
that
\begin{align}
  \label{eq:uv-C} 
  u_{ij}{}^{IJ}(t)=&\, \cosh^3 (2t) \,\delta_{ij}{}^{IJ} 
  \pm \tfrac12 \cosh(2t) \sinh^2 (2t)\, C^\pm_{ijIJ}\,, 
  \nonumber\\[.2ex]
  v_{ij IJ}(t)=&\, \pm \alpha\,\sinh^3 (2t) \, \delta_{ij}{}^{IJ} +
  \tfrac{1}2\alpha \, \sinh(2t) \cosh^2 (2t) \, C^\pm_{ijIJ}\,. 
\end{align}
With these results one can evaluate the corresponding $T$-tensors
\eqref{eq:def-T-tensor}. A straightforward calculation yields the
following results for the component functions $A_1$ and $A_2$,
\begin{equation}
  \label{eq:A-1-2}
  A_1{}^{ij} = \delta^{ij} \,A(t)\,,\qquad A_{2\,i}{}^{jkl} = A_2(t)\,
  C^\pm{}_i{}^{jkl} \,.
\end{equation}
Note that the parameter $t$ parametrizes the vacuum expectation value
of either a selfdual or an anti-selfdual field. We will not consider
both vacuum-expectation values simultaneously for reasons of
simplicity. 
%; it will be obvious from the text which case we are considering. 
When allowing both vacuum-expectation values simultaneously, this
would define a $\mathrm{G}_2$ invariant background, as $\mathrm{G}_2 =
\mathrm{SO}^+(7)\cap\mathrm{SO}^-(7)$. For the special configurations
defined by \eqref{eq:u-v-t} the potential takes the simple form,
\begin{equation}
  \label{eq:reduced-pot}
  \mathcal{P}(t) = g^2 \big[-6\, \vert A_1(t)\vert^2 + 14\,\vert
  A_2(t)\vert^2 \big]\,. 
\end{equation}
Its stationary points are determined by the condition that $\alpha
A_2(t)\big(A_1(t) + 3\, A_2(t)\big)$ is imaginary. 

Making use of \eqref{eq:uv-C} and inserting the deformation parameter
$\omega$ according to \eqref{eq:uv-redef}, leads to the following
expressions for the two functions $A_1(t)$ 
and $A_2(t)$ defined in \eqref{eq:A-1-2},
\begin{align}
  \label{eq:A-12-combined}
  A_1(\omega,t) =&\,  \mathrm{e}^{-\mathrm{i}\omega} \big[ c^7 + 7 c^3  s^4\big] \pm 
  \alpha^\ast\mathrm{e}^{\mathrm{i}\omega} \big[ s^7 + 7 c^4
  s^3\big]\,, \nonumber\\[.2ex] 
  A_2(\omega,t) =&\, \mp\mathrm{e}^{-\mathrm{i}\omega} \big[c\,s^6
  +4 c^3s^4 + 3 c^5 s^2\big] -
  \alpha^\ast\mathrm{e}^{\mathrm{i}\omega} \big[c^6s + 
   4 c^4s^3 + 3 c^2 s^5\big] \,,
\end{align}  
where $c\equiv \cosh(2t)$ and $s\equiv \sinh(2t)$. It is convenient to
present these results as follows.  For the
$\mathrm{SO}(7)^+$-invariant background, we obtain
\begin{align}
  \label{eq:A-12-plus}
  A^+_1(\omega,t) =&\,  \mathrm{e}^{-\mathrm{i}\omega} \big[ c^7 + 7 c^3
  s^4\big] + 
  \mathrm{e}^{\mathrm{i}\omega} \big[ s^7 + 7 c^4
  s^3\big]\,, \nonumber\\[.2ex]
  A^+_2(\omega,t) =&\, -\mathrm{e}^{-\mathrm{i}\omega} \big[c\,s^6
  +4 c^3s^4 + 3 c^5 s^2\big] 
   -\mathrm{e}^{\mathrm{i}\omega} \big[c^6s +
  4 c^4s^3 + 3 c^2 s^5\big] \,,
\end{align}  
whereas for the $\mathrm{SO}(7)^-$-invariant background we write
\begin{align}
  \label{eq:A-12-minus}
  A^-_1(\omega,t) =&\,
  \mathrm{e}^{\mathrm{i}\pi/4}\left\{\mathrm{e}^{-\mathrm{i}\tilde\omega}
    \big[ c^7 +7 c^3 s^4\big] + 
  \mathrm{e}^{\mathrm{i}\tilde\omega} \big[ s^7 +7 c^4
  s^3\big]\right\} \,, \nonumber\\[.2ex]
  A^-_2(\omega,t) =&\, - \mathrm{e}^{\mathrm{i}\pi/4}\left\{- 
   \mathrm{e}^{-\mathrm{i}\tilde\omega} \big[c\,s^6 
  +4 c^3s^4 + 3 c^5 s^2\big] 
  - \mathrm{e}^{\mathrm{i}\tilde\omega} \big[c^6s +
  4 c^4s^3 +3 c^2 s^5\big]\right\} \,, 
\end{align}  
with $\tilde\omega= \omega+\pi/4$. 

Interestingly the two $\mathrm{SO}^\pm(7)$ backgrounds lead to the
same expression for the $T$-tensor, up to an overall phase factor and
a shift in $\omega$, although the overall phase factors for
$A^\pm_1(\omega,t)$ and $A^\pm_2(\omega,(t)$ are clearly not the
same. Because of this relation the two expressions
\eqref{eq:A-12-plus} and \eqref{eq:A-12-minus} enable us to write the
same formula for both potentials, but in terms of different
parameters,
\begin{align}
  \label{eq:potential-pm}
  \mathcal{P}^+ (\omega,t_+) =\frac{g^2}8 \Big\{\cos^2\omega \big(x_+^{14} -
  14 x_+^6 - 35 x_+^{-2}\big) + \sin^2\omega \big(x_+^{-14} - 14
  x_+^{-6} - 35 x_+^2\big)\Big\}\,,
  \nonumber  \\
  \mathcal{P}^- (\omega,t_-) =\frac{g^2}8 \Big\{\cos^2\tilde\omega
  \big(x_-^{14} - 14 x_-^6 - 35 x_-^{-2}\big) + \sin^2\tilde\omega
  \big(x_-^{-14} - 14 x_-^{-6} - 35 x_-^2\big)\Big\}\,,
\end{align}
with $x_\pm\equiv e^{2t_\pm}$.  For $\omega=0$ and $\omega = \pi/2$,
respectively, these formulas reproduce the results of
\cite{deWit:1983gs}; in particular, for $\tilde\omega=\pi/4$ we
re-obtain the $\mathrm{SO}(7)^-$ potential,
\begin{equation}
  \mathcal{P}^-(t_-) = - 2 g^2 \cosh^5(4t_-)\, \big[ 5 - 2\cosh(8t_-)\big]\,. 
\end{equation}

Let us first briefly discuss the stationary points of
$\mathcal{P}_\pm$, suppressing the distinction between the parameters,
$x_\pm$ and between $\omega$ and $\tilde\omega$. Defining $z\equiv x^4
= e^{8t}\geq0$, the condition for the potentials to be stationary
\eqref{eq:potential-pm} is
\begin{equation}
  \label{eq:stat-points}
  (z^2 -1) \left[ \cos^2 \omega \, z^3 (z^2 -5) -  \sin^2\omega
    \left( 5\,z^2 - 1\right)\right] = 0 \,. 
\end{equation}
As it turns out this equation has three solutions. One is $z=1
\Leftrightarrow t=0$. A second solution exists with $0\leq z\leq
1/\sqrt{5}$ and a third one with $\sqrt{5} \leq z$.  When
$\sin\omega=0$, there is a regular solution with $z=\sqrt{5}$
as well as a `run-away solution' $z=0\Leftrightarrow t=-\infty$;
the corresponding solutions for $\cos\omega =0$ are obtained
by interchanging $z\leftrightarrow z^{-1}$ or $t\leftrightarrow -t$.
%The remaining solution then corresponds to $z=\sqrt{5}$ and $z=1/\sqrt{5}$.
For the $\mathrm{SO}(7)^\pm$ solutions, we see that there is
only a single $\mathrm{SO}(7)^+$ solution, $z_+=\sqrt{5}$ or
$z_+=1/\sqrt{5}$ (as already explained above) when $\omega= 0$ and
$\omega = \pi/2$, respectively.  For the $\mathrm{SO}(7)^-$
backgrounds we recover the two solutions at $\tilde\omega=\pi/4$
(corresponding to $\omega=0$ with $\coth 4t_- = \pm \sqrt{5}$. These
two solutions are related by parity reversal. For $\tilde\omega= 0$ or
$\tilde\omega= \pi/2$, there is again a run-away solution.

Let us now examine the consequences of the various equivalences
between different $\omega$-values noted in section
\ref{sec:so(8)-gaugings}. First of all, under a shift $\omega\to
\omega+\pi/2$ the potentials \eqref{eq:potential-pm} change according
to $\mathcal{P}^\pm(\omega+\pi/2,t)= \mathcal{P}^\pm(\omega,-t)$,
which is in agreement with what was derived more generally in section
\ref{sec:so(8)-gaugings}. Furthermore the functions $A^\pm_{1,2}$
satisfy $A^\pm_{1,2}(\omega+\pi/2,t)= - \mathrm{e}^{\mathrm{i}\pi/2}
\,A^\pm_{1,2}(\omega,-t)$ which is consistent with \eqref{eq:phase-T}.
Under the other equivalence associated with the reflection $\omega\to
-\omega$ the two potentials change according to
$\mathcal{P}^\pm(-\omega, t)= \mathcal{P}^\pm(\omega,\pm t)$, which
reflects the fact that for $\omega=0$, $t_+$ is a scalar and $t_-$ is
a pseudoscalar.

It is rather obvious that the {\it separate} potentials
$\mathcal{P}^\pm$ will exhibit no other equivalence relations, and in
particular no relation associated with the shift $\omega\to
\omega+\pi/4$. Indeed, this equivalence is qualitatively different
because it also involves a change of basis for $\mathrm{SO}(8)$, as is
shown in \eqref{eq:redef-A}. Therefore the two potentials are
interchanged!  Inspection shows that the actual relation is given by
\begin{equation}
  \label{eq:2pi-4-change}
  \mathcal{P}^+(\omega+\pi/4, t) =\mathcal{P}^-(\omega, t) \,,
  \qquad
  \mathcal{P}^-(\omega+\pi/4, t) =\mathcal{P}^+(\omega, - t) \,.
\end{equation}
Before explaining this relation in more detail, we note that by applying
this change twice, one recovers the result noted above for the shift
$\omega\to \omega+\pi/2$. 

Let us now clarify the details associated with the equivalence shift
$\omega\to \omega+\pi/4$.  In the new $\mathrm{SO}(8)$ basis the
duality assignments of the $\mathrm{SO}(7)$ invariant tensors change
according to
\begin{equation}
  \label{eq:duality-flip}
  C^\pm_{IJKL} \longrightarrow - C^\mp_{IJKL}\,,
\end{equation}
The change in the duality phase is due to the fact that $\det[P_8]=-1$ so that the
8-dimensional Levi-Civita symbol changes sign. Furthermore, the overall sign in
\eqref{eq:duality-flip} is required in order to re-establish the
normalization condition \eqref{eq:CC}. Using the correspondence noted
below \eqref{eq:CC}, which explains that $\phi^{ijkl} = 2\sqrt{2}
\,\big(-t_+\,C^{+ijkl} +\mathrm{i}t_- \,C^{-ijkl}\big)$, we note the relation,
\begin{equation}
  \label{eq:phi-t-pm}
    \mathrm{e}^{\mathrm{i}\pi/2}\, P_8{}^i{}_m \, P_8{}^j{}_n \, P_8{}^k{}_p
  \, P_8{}^l{}_q  \,\phi^{mnpq} = 2\sqrt{2}\big (t_-\, C^{+ijkl}
  +\mathrm{i} t_+\,C^{-ijkl}\big)\,, 
\end{equation}
where we made use of \eqref{eq:duality-flip}.  With this result we can
evaluate \eqref{eq:phase-T-eq-pi4}, which leads to the following
result,
\begin{align}
  \label{eq:T-transf-quarter}
  T_i{}^{jkl} \big(\omega+\pi/4\,;\, t_+=t,\, t_-=0\big) =&\,
  \mathrm{e}^{-\mathrm{i}\pi/4}\, T_i{}^{jkl} \big(\omega\,;\,
  t_+=0,\, t_-=t\big)
  \,,\nonumber\\[.4ex]
  T_i{}^{jkl} \big(\omega+\pi/4\,; \,t_+=0,\, t_-=t\big) =&\,
  \mathrm{e}^{-\mathrm{i}\pi/4}\, T_i{}^{jkl} \big(\omega\,;\,
  t_+=-t,\, t_-=0\big) \,.
\end{align}
This result is in line with \eqref{eq:2pi-4-change} and can also be
verified explicity on the functions $A_{1,2}^\pm$ shown in
\eqref{eq:A-12-plus} and \eqref{eq:A-12-minus}. One then observes that
$A_2$ acquires an extra minus sign, which is due to the fact that in
the $T$-tensor, $A_2$ is multiplied by the tensor $C^\pm$
(cf. \ref{eq:A-1-2}).

Hence we have explicitly verified all the equivalence relations for
the $\mathrm{SO}(7)^\pm$ solutions. While it is clear that the
equivalence based on the shift $\omega\to \omega+\pi/4$ is more
subtle, these subtleties have been fully accounted for. Our
conclusions are in full agreement with those of
\cite{Dall'Agata:2012bb}. Obviously this pattern will persist for
solutions with less symmetry.
 
%%%%%%%%%%%%%%%%%%%%%%%%%%%%%%%%%%%%%%%%%%%%%%%%%%%%%%%%%%%%%%%%%%%
\section{The embedding in eleven dimensional supergravity}
\label{sec:embedding}
\setcounter{equation}{0}
%%%%%%%%%%%%%%%%%%%%%%%%%%%%%%%%%%%%%%%%%%%%%%%%%%%%%%%%%%%%%%%%%%%%
An important question concerns the possible relation of the deformed
$\mathrm{SO}(8)$ gauged supergravities to $11D$ supergravity as
originally formulated in \cite{Cremmer:1978km}. More specifically, can
the deformed $4D$ supergravities be understood as consistent
truncations of the $11D$ theory? For the undeformed theory this
embedding was studied long ago and it was shown to correspond to a
consistent truncation of $11D$ supergravity \cite{deWit:1986iy};
a particular subtlety related to the $11D$ field strengths
was resolved only recently in \cite{Nicolai:2011cy}.  By a `consistent
embedding' we mean that the full field configuration space of gauged
$N=8$ supergravity can be obtained by consistently truncating $11D$
supergravity, so that all the solutions of the $4D$ theory (including
$x$-dependent ones) can be uplifted to solutions of the
higher-dimensional theory. The original work made use of the
$\mathrm{SL}(8)$ invariant formulation of $N=8$ supergravity, and
therefore our first task is to investigate whether or not the original
approach can be extended to the electric duality basis of the deformed
theories based on \eqref{eq:E-angle}.

We first recall that the consistency proof of
\cite{deWit:1986iy,Nicolai:2011cy} is based on the reformulation of
the $11D$ theory with local $\mathrm{SU}(8)$ invariance that has been
presented in \cite{deWit:1985iy,deWit:1986mz}. This reformulation
relies on a $4+7$ split of the $11D$ theory \cite{Cremmer:1978km}
where the original tangent space group $\mathrm{SO}(1,10)$ is replaced
by $\mathrm{SO}(1,3)\times\mathrm{SU}(8)$, so that the $4D$ R-symmetry
group is realized on the full $11D$ supergravity. In this construction
various features associated with $\mathrm{E}_{7(7)}$ emerge, although
$\mathrm{E}_{7(7)}$ is not a symmetry group of the theory. A key
ingredient in that construction was the so-called {\em generalized
  vielbein}, which is a soldering form defined by
\begin{equation}
  \label{eq:emAB}
  e^m{}_{AB} (x,y) = \mathrm{i} \Delta^{-1/2}\,e_a{}^m  \,\big(
  \Phi^\mathrm{T}\Gamma^a \Phi\big)_{AB} 
  \;\; , \qquad e^{mAB} \equiv (e^m{}_{AB})^*
\end{equation}
where these quantities depend on all eleven coordinates $z^M
\equiv (x^\mu , y^m)$.  Here $e_m{}^a$ is the internal siebenbein that
is part of the elfbein of $11D$ supergravity in a triangular gauge
adapted to the 4+7 split of space-time, 
\begin{equation}
  \label{eq:elfbein}
  E_M{}^A(x,y) = \begin{pmatrix} \Delta^{-1/2} e_\mu{}^\alpha & B_\mu{}^m
    \,e_m{}^a\\[4mm]
    0 & e_m{}^a 
    \end{pmatrix} \;,
\end{equation}
where $\Delta \equiv \det[ e_m{}^a]$ is the metric determinant for the
compact internal space. Tangent-space indices have been denoted by
$\alpha$ and $a$, respectively. Appendix
\ref{App:derivation-dual-gen-vielbein} contains some of the
definitions for the gamma matrices and the spinor fields. The indices
$A,B,\ldots=1,2,\ldots,8$ are initially $\mathrm{Spin}(7)$ indices
associated with the spinor indices of the fermions and the gamma
matrices, but they are elevated to chiral $\mathrm{SU}(8)$ in the
reformulation of the theory. This is achieved by means of the matrix
$\Phi(x,y)\in\mathrm{SU}(8)$ which is required to rewrite the theory
into $\mathrm{SU}(8)$ covariant form.  While this matrix is thus
undetermined prior to truncation, its precise form will be fixed in a
specific truncation modulo the residual ($x$-dependent) local
$\mathrm{SU}(8)$ symmetry of the $N=8$ theory. The underlying idea
here is that the resulting $4D$ spinors can in principle transform
under the $\mathrm{SU}(8)$ R-symmetry, although only the
$\mathrm{Spin}(7)$ subgroup is initially realized as a local
symmetry. Introducing the compensating phase $\Phi$ generalizes the
local symmetry to the full R-symmetry group. To make this approach
viable, it is required that the bosonic quantities that appear in the
supersymmetry transformations of the fermions, constitute
$\mathrm{SU}(8)$ representations.

Subsequently, consider the supersymmetry transformations as they
emerge for the components of the $11D$ metric, evaluated in the
context of the standard Kaluza-Klein decompositions
\cite{deWit:1986mz}, 
\begin{align}
  \label{eq:metric-KK}
  \delta e_\mu {}^\alpha =&\, \tfrac12
  \bar\epsilon^A\gamma^\alpha\psi_{\mu A} + \mathrm{h.c.}\,, \nonumber\\[1mm]
  \delta B_\mu{}^m =&\,\tfrac18\sqrt{2} \,e^m{}_{AB} \,\big(2\sqrt{2}
  \,\bar\epsilon^A \psi_\mu{}^B + \bar\epsilon_C \gamma_\mu \chi^{ABC}
  \big) + \mathrm{h.c.}\,, \nonumber\\[1mm]
  \delta e^m{}_{AB} =&\, -\sqrt{2} \,\Sigma_{ABCD} \,e^{m CD} \,,
\end{align}
where 
\begin{equation}
  \label{eq:def-Sigma}
  \Sigma_{ABCD} \equiv \bar\epsilon_{[A} \chi_{BCD]} +\tfrac1{24}
  \varepsilon_{ABCDEFG} \,\bar\epsilon^{E}\chi^{FGH} \,. 
\end{equation}
We stress that at this point the various quantities all depend on the
coordinates $x^\mu$ and $y^m$. The fermions have been rewritten
according to the same standard Kaluza-Klein procedure; in particular, 
the spin-$\tfrac12$ fields $\chi_{ABC}$ are the chiral components of the
56 fermions that emerge from the $11D$ gravitino fields $\Psi_a$,
see (\ref{eq:spinor-def}) for the precise definitions.

To truncate the $11D$ fields to the $4D$ fields the dependence on the
extra seven coordinates $y^m$ is extracted in the form of the Killing
vectors and Killing spinors of $S^7$ such as to make contact with the
round sphere of a given radius. Then the deviations of the fields away
from the $S^7$ solution are encoded in terms of the $x$-dependent
fields of $4D$ $\mathrm{SO}(8)$-gauged maximal supergravity. The
spinors and vierbein fields can be expressed in the corresponding
quantities of the $4D$ maximal supergravity by exploiting $S^7$
Killing spinors $\eta_A{}^i(y)$ with $i=1,2,\ldots,8$ and their
inverses obeying $\eta_i{}^A\, \eta_B{}^i = \delta^A{\!}_B$ (note that
the fermionic quantities on the left-hand side have all been supplied
with the appropriate compensating SU(8) rotation $\Phi$),
\begin{align}
  \label{eq:vierbein-spinor-trunc}
  \psi_{\mu A}(x,y)=&\,\psi_{\mu i}(x)\, \eta_A{}^i(y)\,,   \nonumber\\
  \chi_{ABC} (x,y)=&\, \chi_{ijk}(x)\,
  \eta_A{}^i(y)\,\eta_B{}^j(y)\,\eta_C{}^k(y)\,, \nonumber\\ 
  e_\mu{}^\alpha(x,y) =&\, e_\mu{}^\alpha(x)\,, \nonumber\\
  \epsilon_A(x,y) =&\,  \epsilon_i(x)\,\eta_A{}^i(y)\,,  \nonumber\\
  U(x,y)^A{}_B =&\, U(x)^i{}_j\, \eta_A{}^i(y)\,\eta^B{}_j(y)\,, 
\end{align}
where $U(x,y)^A{}_B$ is the $\mathrm{SU}(8)$ transformation matrix of
the full $11D$ theory written in the formulation of
\cite{deWit:1986mz}, whereas $U(x)^i{}_j$ is the corresponding matrix
in the $4D$ theory. 
 
In accordance with the standard Kaluza-Klein ansatz,
the vector gauge fields $B_\mu{}^m$ are assumed to be
proportional to the 28 $S^7$ Killing vectors $K^{mIJ}(y)$, labeled by
the 28 antisymmetric index pairs $[IJ]$ (with $I,J=1,2,\ldots,8$),  and
related to the Killing spinors by
\begin{equation}
  \label{eq:K-vectors}
  K^{mIJ} = \mathrm{i} {\stackrel{\circ}{e}}_a{}^{\!\!m}\, \eta^I{}_A\,
  \Gamma^{aAB} \, \eta^J{}_B \,,
\end{equation}
where ${\stackrel{\circ}{e}}_a{}^{\!\!m}(y)$ is the $S^7$ background
siebenbein, so that
\begin{equation}
  \label{eq:B-mu}
  B_\mu{}^m(x,y) = -\tfrac14\sqrt{2}\, K^{mIJ} (y)\,
  A_\mu{}^{IJ}(x)\,. 
\end{equation}
Defining as before, 
\begin{equation}
  \label{eq:emij}
  e^m{}_{ij}(x,y) \equiv e^m{}_{AB} (x,y)\,\eta_i{}^A(y) \,\eta_j{}^B(y) 
  \; , \qquad   e^{m\,ij} \equiv (e^m{}_{ij})^*\,, 
\end{equation}
it follows that $B_\mu{}^m$ and $e^m{}_{ij}$ must have the same
$y$-dependence. Comparing with the $4D$ result from
\cite{deWit:1982ig} for the variation of the 28 electric
vectors,\footnote{%%%%%%%%%%%%%%%%%%%%
  We rescaled the $4D$ supersymmetry parameter $\epsilon$ used in {\it
    e.g.}  \cite{deWit:1982ig,deWit:2007mt} with a factor $\tfrac12$
  in order to be consistent with the $11D$ definitions in
  \cite{deWit:1986mz}.  
} %%%%%%%%%%%%%%%%%%%%%%%%%%%%%%%%
\begin{equation}
  \label{eq:delA}
  \delta A_\mu{}^{IJ} = - \tfrac12 \big(u_{ij}{}^{IJ} + v_{ijIJ} \big) 
  \big( \bar\epsilon_k\gamma_\mu \chi^{ijk} + 2\sqrt{2}\bar\epsilon^i \psi_\mu^j \big) 
  \, + \, {\rm h.c.} \,, 
\end{equation}
one infers the following ansatz for the generalized vielbein, 
\begin{align}
  \label{eq:GV}
  e^m{}_{ij}(x,y) =&\, K^{mIJ}(y)\, \big[ u_{ij}{}^{IJ}(x) +
  v_{ijIJ}(x)\big]\,, \nonumber\\[1mm] 
  e^{m\,ij}(x,y) =&\, K^{mIJ}(y) \,\big[ u^{ij}{}_{IJ}(x) +
  v^{ijIJ}(x)\big]\,,
\end{align}
where $u_{ij}{}^{IJ}$ and $v_{ijIJ}$ are defined by the 56-bein
$\mathcal{V}$ of the $4D$ theory given in \eqref{eq:u-v-phi}. With
these definitions the reader can easily verify that the $y$-dependence
assigned to both sides of the supersymmetry transformations
\eqref{eq:metric-KK}, is consistent.

Let us comment on the above results. First of all, it is remarkable
that the transformations \eqref{eq:metric-KK}, although still based to
the full $11D$ theory, reflect already the structure of the known $4D$
results. Undoubtedly the underlying reason for this is that the
results were written in a form in which the invariance under the local
R-symmetry group $\mathrm{SU}(8)$ of the maximal $4D$ theory is
manifest. This was, of course, an important motivation for following
the approach initiated in \cite{deWit:1986mz}. Another point is that
the structure exhibited in \eqref{eq:metric-KK} is also present in the
general gaugings of $N=8$ supergravity by means of the embedding
tensor approach \cite{deWit:2007mt}. We will return to this aspect in
due course.

Another aspect that deserves attention concerns the way in which the
sub-matrices $u_{ij}{}^{IJ}$ and $v_{ijIJ}$ appear in the ansatz
\eqref{eq:GV} for the generalized vielbein.  But, as we already
explained in section \ref{sec:so(8)-gaugings}, there are alternative
possiblities by changing the electric-magnetic duality frame. For
instance, the electric frame for the new $\mathrm{SO}(8)$ gaugings
requires a different linear combination, namely
$\mathrm{e}^{\mathrm{i}\omega} u_{ij}{}^{IJ}
+\mathrm{e}^{-\mathrm{i}\omega} v_{ijIJ}$, as is indicated by
\eqref{eq:uv-redef}. As we will now argue, it is, however, no longer
possible to have a consistent ansatz with this linear combination,
unless $\exp[2\mathrm{i}\omega]$ is real, so that $\omega$ must be
equal to an integer times $\pi/2$. This implies that the embedding of
the $4D$ fields into the fields of $11D$ supergravity, according to
the scheme followed in \cite{deWit:1986iy}, can only be defined
provided the $4D$ theory is formulated in an $\mathrm{SL}(8)$
invariant duality frame.  The underlying reason for this restriction
is related to the fact that the generalized vielbein $e^m{}_{ij}$, by
virtue of its $11D$ origin \eqref{eq:emAB}, must obey the `Clifford
property',
\begin{equation}
  \label{eq:Clifford}
  e^m{}_{ik}\, e^{n\,kj} + e^n{}_{ik}\, e^{m\,kj} = \tfrac14
  \delta_i{}^j\, e^m{}_{kl} \,e^{n\, lk} \,.
\end{equation}
As  shown in \cite{deWit:1986iy}, \eqref{eq:Clifford} is indeed satisfied with 
\eqref{eq:GV} as a consequence of the properties of the $\mathrm{E}_{7(7)}$ 
matrix  $\cal{V}$ and its submatrices $u$ and $v$. For non-vanishing 
angle $\omega$, the obvious generalization of the formula \eqref{eq:GV} 
would read
\begin{equation}
  \label{eq:modified-GV}
  e^m{}_{ij}(\omega; x,y) = K^{mIJ}(y)\, \big[\mathrm{e}^{\mathrm{i}\omega}
  u_{ij}{}^{IJ}(x) + \mathrm{e}^{-\mathrm{i}\omega} v_{ijIJ}(x)\big]  
\end{equation}
together with its complex conjugate.  However, substituting this
$\omega$-dependent ansatz for the vielbein into \eqref{eq:Clifford},
it turns out that this relation no longer holds for arbitrary values
of $\omega$. To see why this is the case, let us for instance
reconsider equation (2.21) of \cite{deWit:1986iy} and the subsequent
equations. There (\ref{eq:Clifford}) is proven by showing that
$e^{(m}{}_{kl}\, e^{n) ij}$ vanishes upon contraction with an
anti-hermitean traceless $\mathrm{SU}(8)$ matrix $\Lambda_{ij}{}^{kl}
= \Lambda_{[i}{}{}^{[k} \delta_{j]}{}^{l]}$ where $\Lambda^i{}_j = -
\Lambda_j{}^i$ and $\Lambda_i{}^i=0$. Inserting the modified ansatz
\eqref{eq:modified-GV} into the left-hand side of (\ref{eq:Clifford})
the part of the argument involving the $\omega$-independent
combination $u\Lambda \bar{u} + v\Lambda\bar{v}$ goes through as
before. By contrast, the second part of the argument involves the
replacement
\begin{align}
  \label{eq:uLv}
  & \big[ (u\Lambda\bar{v})^{IJ,KL} + (v\Lambda\bar{u})_{IJ,KL} \big]
     \big(K^{m\,IJ} K^{n\, KL}\, + \, K^{m\,IJ} K^{n\, KL}\big) \;
     \rightarrow \nonumber\\[1mm] 
     &\qquad \rightarrow\; \Big[ e^{2\mathrm{i}\omega} (u\Lambda\bar{v})^{IJ,KL}  
  + e^{-2\mathrm{i}\omega} (v\Lambda\bar{u})_{IJ,KL} \Big]
     \big(K^{m\,IJ} K^{n\, KL}\, + \, K^{m\,IJ} K^{n\, KL}\big) \,. 
\end{align}
While for the first line, one could exploit the complex selfduality of
both terms together with the anti-hermiticity of the matrix $\Lambda$
to show that these terms cancel, this argument fails, however, in
presence of the non-trivial phase factor in the second line, even
though the supersymmetry variations based on
\eqref{eq:vierbein-spinor-trunc}, \eqref{eq:B-mu} and
\eqref{eq:modified-GV} do remain mutually consistent (provided that
ones uses the $4D$ transformations in the corresponding
$\omega$-dependent electric-magnetic duality frame). The breaking of
U(8) to its subgroup SU(8) through the presence of the
$\varepsilon$-tensor also vitiates other parts of the proof in
\cite{deWit:1986iy}: in fact, all arguments relying on selfduality or
anti-selfduality (e.g. in the later equations (5.11) and (5.25)) fail
for $\omega \neq 0, \pi/2$ for precisely this reason.

The conclusion is therefore that the embedding of $\omega$-deformed
$\mathrm{SO}(8)$ gaugings into $11D$ supergravity has to be effected
based on the $4D$ theory written in the $\mathrm{SL}(8)$ covariant
formulation. This implies that one has to deal with an
electric-magnetic duality frame that is not purely electric, while the
concept of magnetic charges does not exist in the context of eleven
dimensions. The formulation of the $4D$ theory that accomplishes this
in four dimensions, is the embedding-tensor formulation of maximal
$N=8$ supergravity given in \cite{deWit:2007mt}. In this approach all
the couplings of the ungauged theory retain their original form given
in \cite{deWit:1982ig}, but the $\mathrm{SO}(8)$ generators will
change and will involve magnetic components. In the embedding tensor
formalism there are also magnetic gauge fields that couple to these
magnetic components, but at the same time there are additional tensor
fields with certain gauge invariances and constraints that ensure that
28 linear combinations of the electric and magnetic gauge fields are
suppressed. Therefore only 28 gauge fields remain which will
correspond to $\omega$-dependent linear combinations of the original
28 electric and 28 magnetic gauge fields. A natural question is
therefore whether some of the ingredients of the embedding tensor
formalism will also play a role in this context and reveal how the
magnetic sector of the $4D$ theory can emerge in a possible embedding
in $11D$ supergravity for arbitrary values of $\omega$.

Let us therefore further clarify some details of the $4D$ embedding
tensor approach in the $\mathrm{SL}(8)$ frame. Casting the results of
\cite{deWit:2007mt} in this frame shows that the electric and magnetic
gauge fields transform under supersymmetry as,
\begin{align}
  \label{eq:e-m-gauge-fields-4D}
  \delta A_\mu{}^{IJ} =&\, - \tfrac12 \big(u_{ij}{}^{IJ} + v_{ijIJ} \big) \big(
  \bar\epsilon_k\gamma_\mu \chi^{ijk} + 2\sqrt{2}\,\bar\epsilon^i
  \psi_\mu{}^j \big) \, + \, {\rm h.c.}\,, \nonumber\\[.2ex]
  \delta {A}_{\mu IJ} =&- \tfrac12\mathrm{i} \big(u_{ij}{}^{IJ} - v_{ijIJ}
  \big) \big( \bar\epsilon_k\gamma_\mu \chi^{ijk} +
  2\sqrt{2}\,\bar\epsilon^i \psi_\mu{}^j \big) \, + \, {\rm h.c.}\,.
\end{align}
Obviously the identification of these `magnetic' gauge fields in $11D$
supergravity should be a crucial element in establishing a possible
$11D$ origin of the $\omega$-deformed theories.

Another aspect concerns the relation between the $4D$ $T$-tensors and
the $11D$ theory. In $4D$ the $T$-tensor is generated by the embedding
tensor that defines how the 56 gauge fields couple to the generators
of the group $\mathrm{E}_{7(7)}$, and therefore to the electric and
the magnetic generators. The latter generate composite electric
`connections' $\mathcal{B}_{IJ}$ and $\mathcal{A}_{IJ}$, belonging to
the $\bf{63}$ and $\bf{70}$ representations of $\mathrm{SU}(8)$, which
together comprise the $\bf{133}$ representation of
$\mathrm{E}_{7(7)}$. Likewise there are also magnetic `connections'
$\mathcal{B}^{IJ}$ and
$\mathcal{A}^{IJ}$.~\footnote{%%%%%%%%%%%%%%%%%%%%%%%%%%%%%%%%%%
  In \cite{deWit:2007mt} these `connections' were denoted by
  $\mathcal{Q}_M$ and $\mathcal{P}_M$, where the index $M$ is an index
  belonging to the $\boldsymbol{56}$ representation of
  $\mathrm{E}_{7(7)}$, which decomposes into the electric and magnetic
  28-dimensional representations of $\mathrm{SL}(8)$.
} %%%%%%%%%%%%%%%%%%%%%%%%%%%%%%%%%%%%%%%%%%%%%%%%%%%%%%%
Obviously these connections do not constitute vectors in some
underlying continuous space, but nevertheless they are the
straightforward generalization of the space-time connections
$\mathcal{B}_\mu$ and $\mathcal{A}_\mu$ that are already present in
the ungauged supergravity. For instance, the $\mathcal{B}_\mu$ provide
the composite gauge fields for the $\mathrm{SU}(8)$ gauge group.

In the locally $\mathrm{SU}(8)$ invariant formulation of $11D$
supergravity, there is a similar situation, namely there exist
connections $\mathcal{B}_M$ and $\mathcal{A}_M$, but now these are
vector fields in the $11D$ space-time, decomposing into the $4D$
vectors $\mathcal{B}_\mu$ and $\mathcal{A}_\mu$, and the $7D$ vectors
$\mathcal{B}_m$ and $\mathcal{A}_m$. These connections are present in
the supersymmetry transformations of the fermion fields. But they also
emerge as composite $\mathrm{E}_{7(7)}$ connections in the so-called
{\it generalized vielbein postulate}, which expresses the fact that
the generalized vielbein is covariantly constant \cite{deWit:1986mz}.
Obviously the connections $\mathcal{B}_m$ and $\mathcal{A}_m$ are
expected to be related to the analogous `connections'
$\mathcal{B}_{IJ}$ and $\mathcal{A}_{IJ}$ (and possibly their magnetic
duals) that appear in the $4D$
theories. Indeed, this expectation is precisely confirmed for the case
of the original $\omega=0$ supergravity as was exhibited in
\cite{deWit:1986iy}, where the connections $\mathcal{B}_m$ and
$\mathcal{A}_m$ yield the electric $T$-tensor in the $\mathrm{SL}(8)$
duality frame.

%%%%%%%%%%%%%%%%%%%%%%%%%%%%%%%%%%%%%%%%%%%%%%%%%%%%%%%%%
\section{Dual quantities and the non-linear flux ansatz}
\label{sec:dual-quantities}
\setcounter{equation}{0}
%%%%%%%%%%%%%%%%%%%%%%%%%%%%%%%%%%%%%%%%%%%%%%%%%%%%%%%%%
In the past the question whether $11D$ supergravity possesses certain
structures in the context of a lower-dimensional formulation that more
fully exhibits the duality symmetry, has been analyzed in the case of
$3D$ where the duality group equals $\mathrm{E}_{8(8)}$, implying a
kind of `generalized geometry' based on $\mathrm{E}_{8(8)}$
\cite{Koepsell:2000xg}. This effort (as well as more recent efforts in
connection with `generalized geometry') was based on a quest for
further unification, while in the context of this paper one is
confronted with a more concrete motivation, namely of how to reconcile
the deformed $\mathrm{SL}(8)$ supergravities in $4D$ with the full
$11D$ theory. Another main difference is that in $4D$ we have the
possibility of testing the various formulas for non-trivial
compactifications, whereas in $3D$ most gaugings cannot be obtained
from (and thus not compared with) spontaneously compactified solutions
of the $11D$ theory. A rather surprising consequence of the present
analysis is that we are in this way led to a simple formula for the
non-linear flux ansatz!

As was pointed out in the previous section, it is obviously important
in this context to have both electric gauge fields and their dual
magnetic ones. Taking this as a guideline, we are led to ask whether
the $11D$ theory contains such dual gauge fields, and whether those
have a relation to components of the three-form tensor fields
$A_{MNP}$. The latter fields were avoided in the analysis of
\cite{deWit:1986iy}, because the equations of motion and the
supersymmetry variations of $11D$ supergravity only involve the
four-form field strengths, and the truncation to $4D$ usually involves
tensor-scalar dualities which require more detailed knowledge of the
truncated Lagrangian. Furthermore, for the $S^7$ compactification of
$11D$ supergravity all 28 spin-1 degrees of freedom are known to
reside in the Kaluza-Klein vector $B_\mu{}^m$ according to
\eqref{eq:B-mu}.  By contrast, for the toroidal truncation of
\cite{Cremmer:1979up} only seven (electric) spin-1 degrees of freedom
originate from $B_\mu{}^m$, while the remaining 21 (magnetic) spin-1
degrees of freedom reside in $A_{\mu mn}$.

We therefore proceed on the assumption that the dual magnetic gauge 
fields are contained in the fields
\begin{equation}\label{eq:Bmumn}
B_{\mu mn}\equiv A_{\mu mn} - B_\mu{}^p\, A_{pmn} \, ,
\end{equation}
which  follow from the standard Kaluza-Klein ansatz and define covariant
vector fields in $4D$. A somewhat subtle calculation (see
\cite{deWit:1986mz} and appendix) shows that these fields transform 
as follows under supersymmetry, 
\begin{align}
  \label{eq:delta-A-mu-mn}
  \delta B_{\mu mn}=&\, - \mathrm{i}\Delta^{-1/2}\big[ \tfrac1{48} 
  (\Gamma_{mn})_{AB} + \tfrac18 \sqrt{2} \,A_{mnp} \,(\Gamma^p)_{AB}\big]\,
  \big(2 \sqrt{2}\,\epsilon^A\psi_\mu{}^B +\bar\epsilon_C \gamma_\mu
  \chi^{ABC} \big)  &  %\nonumber\\ 
  %&\,
     + \; \mathrm{h.c.} \,  ,
\end{align}
where again all redefinitions required in the passage from $11D$ to
$4D$ must be taken into account. As for \eqref{eq:metric-KK}, this
result still reflects the full $11D$ situation since we have not
imposed any restrictions on the dependence on the internal coordinates
$y^m$. Remarkably, the spinor bilinears that appear in
\eqref{eq:delta-A-mu-mn} are exactly as in $\delta B_\mu{}^m$, as well
as in the $4D$ supersymmetry variations of the electric and magnetic
gauge fields, $\delta A_\mu{}^{IJ}$ and $\delta A_{\mu IJ}$, that
follow from the embedding tensor formalism
(cf. \ref{eq:e-m-gauge-fields-4D}) . This indicates that we are
dealing with a {\em dual generalized vielbein}, in terms of which the
supersymmetry variations of $B_\mu{}^m$ and $B_{\mu mn}$ acquire the
same form,
\begin{align}
  \label{eq:delta-B-B}
    \delta B_\mu{}^m =&\,\tfrac18\sqrt{2} \,e^m{}_{AB} \,\big(2\sqrt{2}
  \,\bar\epsilon^A \psi_\mu{}^B + \bar\epsilon_C \gamma_\mu \chi^{ABC}
  \big) \; + \; \mathrm{h.c.}\,, \nonumber\\[1mm]
      \delta B_{\mu mn} =&\,\tfrac18\sqrt{2} \,e_{mn\, AB} \,\big(2\sqrt{2}
  \,\bar\epsilon^A \psi_\mu{}^B + \bar\epsilon_C \gamma_\mu \chi^{ABC}
  \big) \;  + \; \mathrm{h.c.}\,.
\end{align}
Here the normalization of $e_{mn\,AB}$ has been chosen such that the
factors on the right-hand side of the above two equations are
equal. The generalized vielbein \eqref{eq:emAB} is thus complemented
by the following new vielbein-like object
\begin{align}
  \label{eq:dual-gen-vielbein}
  e_{mn\,AB} =&\, -\tfrac1{12} \mathrm{i} \sqrt{2}\,\Delta^{-1/2}\,\Big[
  e_m{}^a e_n{}^b \, \big(\Phi^\mathrm{T} \Gamma_{ab}\Phi\big){}_{AB}
  +6 \sqrt{2} \,A_{mnp}
    \,\big(\Phi^\mathrm{T}\Gamma^p\Phi\big){}_{AB} \Big]
    \,,\nonumber\\
    e_{mn}{}^{AB} \equiv &\, \big(e_{mn\,AB}\big)^\ast \,,
\end{align}
characterized by a pair of {\em lower} world indices $m,n$.  Note that
this new vielbein is complex even in the special gauge $\Phi =
\oneone$. It remains to determine its supersymmetry variation.  In
analogy with the third equation of \eqref{eq:metric-KK}, which was
originally derived in \cite{deWit:1986mz}, one finds that both
vielbeine transform uniformly,
\begin{align}
  \label{eq:susy-var-dual-vielbein}
  \delta e^m{}_{AB} =&\, -\sqrt{2}\, \Sigma_{ABCD} \, e^{m \,CD}
  \,, \nonumber\\[1mm]
\delta e_{mn\,AB} =&\,  -\sqrt{2}\, \Sigma_{ABCD} \, e_{mn}{}^{CD} \,. 
\end{align}
We relegate a derivation of this result to appendix
\ref{App:derivation-dual-gen-vielbein}, where we also summarize a
number of other relevant definitions. The new vielbein
\eqref{eq:dual-gen-vielbein} and the SU(8) covariant supersymmetry variations
\eqref{eq:susy-var-dual-vielbein} are in precise analogy with results
found for the $3+8$ split appropriate to $D=3$ dimensions \cite{Koepsell:2000xg}.

Defining
\begin{equation}
  \label{eq:emnij}
  e_{mn\,ij}(x,y) \equiv e_{mn\,AB} (x,y)\,\eta_i{}^A(y) \,\eta_j{}^B(y) 
  \; , \qquad   e_{mn}{}^{ij} \equiv (e_{mn\, ij})^*\,, 
\end{equation}
one can now derive certain relations for products of the generalized
vielbein, in analogy to the Clifford relation
\eqref{eq:Clifford}. The most obvious one is, 
\begin{equation}
  \label{eq:mix-clifford}
  e_{mn\,ij}\, e^{p\,ij} = -8\, \Delta^{-1}\, {g}^{pq} \,A_{mnq}\,,  
\end{equation}
which defines $A_{mnp}$ in terms of the generalized vielbeine. This
formula is the analog of the corresponding formula for the inverse 
densitized metric $\Delta^{-1} g^{mn}$, obtained by tracing the 
Clifford relation \eqref{eq:Clifford}. An important consequence
of that formula was the {\em non-linear metric ansatz}
\cite{deWit:1984nz,deWit:1986iy},  
\begin{equation}
  \label{eq:gmn}
  \Delta^{-1} g^{mn}(x,y) = \tfrac18 K^{mIJ}(y)\, K^{n KL}(y)\, 
  \big[ (u^{ij}{}_{IJ}{} + v^{ij IJ })( u_{ij}{}^{KL}  + v_{ij KL}) \big] (x)\,,
\end{equation}
where we note that explicit symmetrization in the indices $m$ and $n$
is not necessary owing to the properties of the matrices $u$ and $v$.
With the previously derived formulas \eqref{eq:e-m-gauge-fields-4D}
and \eqref{eq:delta-B-B} we can now deduce, in complete analogy with
\eqref{eq:GV}, a similar ansatz for the dual guage field and the dual
vielbein in the truncation of the $11D$ to the $4D$ fields, {\em viz.}
\begin{align}
  \label{eq:dual-vielbein-ansatz}
  B_{\mu mn}(x,y) =&\, -\tfrac14\sqrt{2}\,\lambda\, K_{mn}{}^{IJ}(y) \, A_{\mu
    IJ}(x) \,,   \nonumber\\[.2ex]
  e_{mn\,ij} (x,y) =&\, \mathrm{i}\lambda\, K_{mn}{}^{IJ}(y)
  \,\big[u_{ij}{}^{IJ} -v_{ijIJ} \big] (x) \,,
\end{align}
where $\lambda$ is an undetermined constant and
\begin{equation}
  K_{mn}{}^{IJ}(y)  \equiv  \; {\stackrel{\circ}{e}}{}_{ma}(y)
  \,{\stackrel{\circ}{e}}{}_{nb}(y) \, \eta^I _A(y) \,\Gamma^{ab\,AB}
  \,\eta^J_B(y) \,.
\end{equation}
Using \eqref{eq:mix-clifford}, \eqref{eq:gmn} and
\eqref{eq:dual-vielbein-ansatz} we get~\footnote{ %%%%%%%%%%
  Although $A_{mnp}$ is only determined up to an $(x,y)$-dependent
  tensor gauge transformation, the truncation fixes the $y$-dependence
  so that $A_{mnp}$ is obtained in a particular gauge. }
%%%%%%%%%%%%%%%%%%%%
\begin{align}
  \label{eq:Amnp}
  & \mathrm{i}\lambda\, K_{mn}{}^{IJ}(y)\, K^{p KL}(y)\,
  \big[(u^{ij}{}_{IJ}{}  -  v^{ij IJ})( u_{ij}{}^{KL}  + v_{ij KL}) \big] (x)= \nonumber\\[2mm]
  & \qquad = \; - 
  K^{pIJ}(y)\, K^{q KL}(y) \,
  \big[ (u^{ij}{}_{IJ}{}  + v^{ijIJ })( u_{ij}{}^{KL}  + v_{ij KL})\big](x)\,
   A_{mnq} (x,y)\,, 
\end{align}
where we remember that the curved indices on the Killing vector $K$ and its derivative 
are always to be raised and lowered with the {\em round} $S^7$ metric. Using
properties of the matrices $u$ and $v$ given in \cite{deWit:1982ig} 
%(such as the relation $u_{ij}{}^{IJ} v^{ijKL} = u_{ij}{}^{KL} v^{ijIJ}$) 
this can be rewritten as 
\begin{align}
  \label{eq:A-mnp}
  & \mathrm{i}\lambda\,K_{mn}{}^{IJ} \,K^{p KL} 
  \big[ v^{ijIJ} v_{ijKL} - v_{ijIJ} v^{ijKL} + u_{ij}{}^{IJ} v^{ijKL}
  -u^{ij}{}_{IJ} v_{ijKL}\big]  = \nonumber\\[2mm]
  & \quad =  \Big[8\, 
   \stackrel{\circ}{g}{}^{pq} - K^{p IJ} K^{qKL} \big( v^{ijIJ}
   v_{ijKL} + v_{ijIJ} v^{ijKL} + u_{ij}{}^{IJ} v^{ijKL} +
   u^{ij}{}_{IJ} v_{ijKL}\big)  \Big]\,A_{mnq}  \,. 
\end{align} 
Observe that both sides of this equation are purely imaginary provided
that $A_{mnp}$ is real, which is precisely as expected. Alternatively
the reality can be proven from the fact that $e_{mn\, ij}\, e^{p\, ij}
= e_{mn}{}^{ij}\, e^p{}_{ij}$, which follows by making use of the
properties of the matrices $u$ and $v$. The expressions
\eqref{eq:Amnp} and \eqref{eq:A-mnp} are the analog of the non-linear
metric ansatz \eqref{eq:gmn}, but now for the three-form field
$A_{mnp}(x,y)$ ({\it alias} the `flux field'). The formulae
\eqref{eq:Amnp} and \eqref{eq:A-mnp} are rather similar to the
conjectured formula~(6.2) in \cite{deWit:1984nz}.  Both results
reproduce the same {\em linear} ansatz for $A_{mnp}$. This illustrates
the difficulty in obtaining consistent non-linear ans\"atze: there is
no way of guessing the correct answer from the linearized expression!

To verify that \eqref{eq:dual-vielbein-ansatz}, and hence
\eqref{eq:Amnp} are really correct we perform a number of consistency
checks.  One such check concerns the constraint, 
\begin{equation}
  e_{mn\, ij} \, e^{n\, ij} = 0\,,
\end{equation} 
which follows from \eqref{eq:mix-clifford} and the antisymmetry of $A_{mnp}$.
To prove it we make use of the identity
\begin{equation}
  \label{eq:KmnKn}
  K_{mn}{}^{IJ} K^{n KL} = - 4 \, \delta^{[ J[K} K_m{}^{L] I]} + 
  K_{mn}{}^{[IJ} K^{n KL]} 
\end{equation}
Now we observe that the first two terms in brackets on the left-hand side 
of \eqref{eq:A-mnp} are anti-symmetric under interchange of the index 
pairs $[IJ]$ and $[KL]$, whence for them, only the first term on the right-hand side 
of \eqref{eq:KmnKn} contributes, so the result of the index contraction is proportional to
\begin{equation}
  \label{eq:KK-uu-asymm}
  \delta^{[ J[K} K_m{}^{L] I]} 
  \big[ u_{ij}{}^{IJ} u^{ij}{}_{KL} - v_{ijIJ} v^{ijKL} \big] = 0\,.
 \end{equation}
The vanishing of this expression follows from the fact that, with 
{\em uncontracted} $\mathrm{SU}(8)$ index pairs $[ij]$ and $[kl]$,
%\begin{equation}
$$
\delta^{[ J[K} K_m^{L] I]} 
 \big[ u_{ij}{}^{IJ} u^{kl}{}_{KL} - v_{ijIJ} v^{klKL}\big]
$$
%\end{equation}
must belong to the Lie algebra of $\mathrm{E}_{7(7)}$ and must
therefore vanish when traced with $\delta^{ij}_{kl}$ over the $\mathrm{SU}(8)$
index pairs $[ij]$ and $[kl]$.  The same argument applies to the
remaining two terms in the bracket on the left-hand side of
\eqref{eq:A-mnp} which are each symmetric under the interchange $[IJ]
\leftrightarrow [KL]$, leaving us with
\begin{equation}
  \label{eq:KK-uu-symm}
K_{mn}{}^{[IJ} K^{n KL]} \big[ u_{ij}{}^{IJ} v^{ijKL}  -u^{ij}{}_{IJ}
v_{ijKL}\big]  = 0\,, 
\end{equation}
because  $K_{mn}{}^{[IJ} K^{n KL]}$ is (complex) selfdual. 

A stronger test, which implies the previous one, is to verify the 
complete anti-symmetry of $A_{mnp}$ in the indices $[mnp]$
from the definition \eqref{eq:Amnp}.
Since the anti-symmetry in $[mn]$ is manifest we need only 
ascertain the anti-symmetry with respect to the other index pair 
$[mp]$, or equivalently $[np]$. This is equivalent to checking 
the anti-symmetry of  $( \Delta^{-1} g^{nr})( \Delta^{-1} g^{ps}) A_{mrs}$ 
in the indices $[np]$. Using \eqref{eq:KmnKn} this requires
\begin{eqnarray} 
  \label{eq:lambda-eq}
  && K^{nKL} K^{p PQ} \big( - 4\, \delta ^{K'M} K_m^{L'N} +
  K_m^{K'L'MN} \big) \nonumber\\[2mm] 
  && \quad \times \, \big( u_{kl}{}^{KL} + v_{kl KL}\big)  \big( u^{kl}{}_{K'L'} + v^{kl K'L'}\big)
  \big( u_{ij}{}^{MN} -  v_{ijMN}\big)  \big( u^{ij}{}_{PQ} + v^{ij PQ}\big)
\end{eqnarray}
to be anti-symmetric in $[np]$.  We now invoke the previous argument
to show that the expression involving the $(u+v)(u-v)$ factor in the
middle is $\mathrm{E}_{7(7)}$ Lie-algebra valued in the index pairs
$[ij]$ and $[kl]$ and hence can be written as $\delta^{[k}{\!}_{[i}
\,\Lambda^{l]}{\!}_{j]}$, with $\Lambda^i{}_j$ anti-hermitean and
traceless. Hence we are left with the task to show that
\begin{equation}
  \Lambda^k{}_i \, \times \big( K^{n KL} K^{p PQ} +  K^{p KL} K^{n PQ} \big) 
  \big( u_{kl}{}^{KL} + v_{kl KL}\big)  \big( u^{il}{}_{PQ} + v^{il
    PQ}\big) = 0\,. 
\end{equation}
Now we invoke the $\mathrm{E}_{7(7)}$ Lie algebra once again: upon
symmetrization under $[KL] \leftrightarrow [PQ]$ it follows that
\begin{align}
  \label{eq:lambda3}
  u^{KL}{}_{ik} \Lambda^i{}_j u^{jk}{}_{PQ} \,+\, v_{KLik}
  \Lambda^i{}_j v^{jkPQ} \, \cong& \;
  u^{KL}{}_{ik}  \Lambda^i{}_j u^{jk}{}_{PQ}  \, +  \,v_{PQik}
  \Lambda^i{}_j v^{jkKL}  \nonumber\\[2mm] 
  =&\; u^{KL}{}_{ik} \Lambda^i{}_j u^{jk}{}_{PQ} \, - \, v^{KLik}
  \Lambda_i{}^j v_{jkPQ} \nonumber\\[2mm]
  =&\; \delta^{[K}{\!}_{[P} X^{L]}{}_{Q]}
\end{align}
is Lie-algebra valued in the index pairs $[KL]$ and $[PQ]$ with an
anti-hermitean and traceless matrix $X^I{}_J$. Hence, this contribution is
proportional to
\begin{equation}
  \label{eq:uu-vv}
  \big( K^{nKL} K^{pKQ} + K^{nKL} K^{pKQ}\big) X^K{}_Q 
  \; \propto \; \; \stackrel{\circ}{g}{}^{np} \,X^K{}_K = 0\,. 
\end{equation}
For the remaining two terms we use for the first term that
\begin{align} 
  \label{eq:uv}
  u_{ik}{}^{KL}\Lambda^i{}_j v^{jk PQ} \, + \,
  u_{ik}{}^{PQ}\Lambda^i{}_j v^{jk KL}\, =&\; u_{ik}{}^{KL}\Lambda^i{}_j
  v^{jk PQ} \, - \,
  v^{ik KL}  \Lambda_l{}^i u_{ik}{}^{PQ} \nonumber\\[2mm]
  =&\; \mbox{complex selfdual in $[KLPQ]$}\,.
\end{align}
For the second term we note, 
\begin{equation}
  \label{eq:vu}
  v_{ikKL}\Lambda^i{}_j u^{jk}{}_{ PQ} \, + \, v_{ikPQ}\Lambda^i{}_j u^{jk}{}_{ KL}
  \,= \, - \big( u^{ik}{}_{KL}\Lambda_i{}^j v_{jk PQ} \, - \,v_{ik KL} \Lambda^i{}_j u^{jk}{}_{PQ}) \; ,
\end{equation}
which equals minus the hermitean conjugate of the first term
\eqref{eq:uv}. Hence, after contraction with $K^{n[KL} K^{p\, PQ]}$, the sum of the two
terms gives zero. Therefore $A_{mnp}(x,y)$ as determined from
\eqref{eq:Amnp} is indeed fully anti-symmetric. 

%%%%%%%%%%%%%%%%%%%%%%%%%%%%%%%%%%%%%%%%%%%%%%%%%%%%%%%%%%%%%%%%%%%
\section{Outlook}
\label{sec:Membedding}
\setcounter{equation}{0}
%%%%%%%%%%%%%%%%%%%%%%%%%%%%%%%%%%%%%%%%%%%%%%%%%%%%%%%%%%%%%%%%%%%
The present work opens unexpected new perspectives on $11D$
supergravity, and the link between this theory and the duality
symmetries of $4D$ maximal supergravity. Although the duality between
electric and magnetic vector fields is normally viewed as a phenomenon
strictly tied to four space-time dimensions, our analysis has revealed
$11D$ structures directly associated to electric-magnetic vector
duality, yielding as a by-product the long sought formula for the
non-linear flux ansatz. These new structures appear in the form of a
dual generalized vielbein $e_{mn\,AB}$, whose properties need to be
explored further. For instance there is the question whether this
object obeys a generalized vielbein postulate analogous to the one
satisfied by $e^m{}_{AB}$ \cite{deWit:1986mz}. The fact that the
solution of the vielbein postulate is not unique, but only determined
up to an homogeneous contribution \cite{Nicolai:2011cy} is likewise
expected to play a role here.

The subtleties regarding the emergence of electric {\em vs.}  magnetic
gauge fields have not been explored much in the present Kaluza-Klein
context. Therefore we briefly return to the issue of the origin of the
dual vector fields from $11$ dimensions, and to the question whether
and how the $\omega$-rotation might be implemented in eleven
dimensions.  One important feature here is that the distribution of
the 28 {\it physical} spin-one degrees between electric and magnetic
vectors depends on the compactification. This is very similar to what
happens in four dimensions in the context of the embedding tensor
formalism, where the embedding tensor determines which combination of
the electric and magnetic gauge fields will eventually carry the
physical spin-one degrees of freedom. For the $S^7$ compactification,
all 28 vector fields reside in the Kaluza-Klein vector field
$B_\mu{}^m(x,y)$ and are electric. By contrast, for the torus
reduction of \cite{Cremmer:1979up} there are only seven electric
vectors associated to the seven Killing vectors on $T^7$, while the
remaining 21 vectors come from $A_{\mu mn}$ and are magnetic. For the
$S^7$ compactification, this raises the question how the theory
manages to prevent the massless excitations contained in $A_{\mu mn}$
from appearing as {\em independent} spin-one degrees of freedom on the
mass shell.
 
One may wonder why, now that a number of the appropriate dual
quantities in the $11D$ theory has been identified, it is not possible
to give a more precise scenario of how the $\omega$-deformations might
be embedded. Let us recall that in \cite{deWit:1986iy}, the $T$-tensor
of the $4D$ supergravity followed from the composite connections
$\mathcal{B}_m$ and $\mathcal{A}_m$, which belong to the ${\bf 133}$
representation of $\mathrm{E}_{7(7)}$. The actual expressions for
these connections in the truncation were determined by solving the
generalized vielbein postulate. When going through the actual
derivation in \cite{deWit:1986iy}, it is difficult to envisage a
modification of the solution that would enable one to include the
magnetic duals. On the other hand, as mentioned above, the 
solution of the generalized vielbein postulates is not unique
\cite{Nicolai:2011cy}, a fact that could possibly be explored to
somehow include the magnetic duals. However, it was also noted in
that work that the ambiguities in $\mathcal{B}_m$ and $\mathcal{A}_m$
are such that they will cancel in the final expression for the
$T$-tensor. Clearly, it is still premature to draw any definite
conclusions from this, given the fact that the dual structures have
not been explored extensively so far, but we expect that the further
analysis of these structure, and in particular, of the generalized vielbein 
postulate for the new vielbein may provide valuable hints as to the 
`hiding place' of the embedding tensor in eleven dimensions.

To better understand the possible origin of the full set of 28 vectors
and their 28 magnetic duals {\em from eleven dimensions} it may be
helpful to recall that the $11D$ theory also allows for dual fields,
although these do not appear in the Lagrangian and transformation
rules of \cite{Cremmer:1978km}.  These are the 6-form field
$A_{MNPQRS}$ (dual to the three-form field $A_{MNP}$) and the `dual
graviton' $h_{M | N_1\cdots N_8}$ (which is dual to the linear
graviton field $h_{MN}$; see {\it e.g.} \cite{Bekaert:2002uh} and
references therein). The latter belongs to a non-trivial Young tableau
representation, which is fully antisymmetric in the last eight indices
$N_1\cdots N_8$ and obeys the irreducibility constraint $h_{[M |
  N_1\cdots N_8]} = 0$.  We note here that the incorporation of the
dual graviton has so far been achieved {\em only at the linearized
  level}, and one may therefore anticipate difficulties in
re-formulating the $11D$ theory in a way that would consistently
incorporate these dual fields {\em at the interacting level}, and in a
way maintaining full $11D$ covariance.~\footnote{In fact, it has been
  known for a long time that even the consistent incorporation of the
  dual 6-form field in the Lagrangian encounters problems, although
  this field can be incorporated in the equations of motion
  \cite{Nicolai:1980kb,Bandos:1997gd}.}  Upon dimensional reduction on
a 7-torus these fields give rise to the full set of 28 + 28 vector
fields (cf. eqs. \eqref{eq:elfbein} and \eqref{eq:Bmumn} for the first
two lines)
\begin{align}\label{11Dduals}
E_M{}^A   \quad &\rightarrow  \quad B_\mu{}^m    \qquad &  \in & \quad {\bf 7}
\nonumber\\ 
A_{MNP}   \quad &\rightarrow  \quad B_{\mu\, mn}      & \in & \quad
\overline{\bf 21} \nonumber\\ 
A_{MNPQRS}   \quad &\rightarrow  \quad \widetilde{B}_{\mu\, npqrs}   &
\in  &  \quad {\bf  21} \nonumber\\ 
h_{M|N_1\cdots N_8}   \quad &\rightarrow  \quad  
B_{m | \mu\, n_1\cdots n_7} \equiv \widetilde{B}_{\mu \,m} \,
\varepsilon_{n_1\cdots n_7}       & \in &\quad   \overline{\bf {7}} 
\end{align}
at least in the linearized analysis (note that $B_{m|\mu n_1\cdots n_7}$ 
does satisfy the irreducibility constraint appropriate to the dual
graviton field, because the Latin indices only run over $1,...,7$).
Here we have indicated the SL(7) (or
GL(7)) representation on the right-hand side. These representations can be
re-combined into the proper SL(8) representations of the electric and
magnetic vectors of $N=8$ supergravity in accordance with the
decomposition
\begin{equation}\label{28+28}
{\bf 28} \oplus \overline{\bf 28} \quad \rightarrow \quad 
{\bf 7} \oplus {\bf 21} \oplus \overline {\bf 7} \oplus \overline{\bf 21}
\end{equation}
This is consistent with the fact that the electric and magnetic fields
must transform in conjugate (`dual') representations. However, as we
said, the distribution of the physical spin-one degrees of freedom
between these fields depends on the compactification. Of course, for
the torus reduction the (ungauged) $4D$ theory cannot tell the
difference between `electric' and `magnetic', but the distinction does
become relevant for the gauged theory, as is evident from the
existence of inequivalent $\omega$-deformed SO(8) gaugings
\cite{Dall'Agata:2012bb}, and from our discussion in section~2.

The decomposition \eqref{28+28} suggests that our set of vielbeine
$(e^m{}_{AB} \,,\, e_{mn AB})$ is still incomplete, and that there
should exist a complementary set $(e_{m\, AB}\, , \, e^{mn}{}_{AB})$
of yet another set of 28 vielbein components that would complete the
generalized vielbein to a full 56-bein in $D=11$ dimensions --- this
was, in fact, the conclusion reached in \cite{Koepsell:2000xg} for
$\mathrm{E}_{8(8)}$ and the 3+8 decomposition of $11D$ supergravity.
Accordingly, the supersymmetry transformations \eqref{eq:metric-KK}
would have to generalize to this hypothetical 56-bein, and the vector
transformations \eqref{eq:delA} and \eqref{eq:e-m-gauge-fields-4D}
would likewise have to follow from a single variation in
$11D$. However, in order to derive these relations we would have to
know the full {\em non-linear} $11D$ transformations of the dual
fields in \eqref{11Dduals}! The $\omega$-dependent vielbein ansatz
\eqref{eq:modified-GV} would then simply follow from
\begin{equation}\label{U1}
e^m{}_{ij}( \omega; x,y) = 
\cos\omega \, e^m{}_{ij}(x,y)  \, + \, \sin\omega \, e_{m\, ij}(x,y)
\end{equation}
thus involving a $\mathrm{U}(1)$ rotation between the Kaluza-Klein
vector $B_\mu{}^m$ and the dual graviton vector $\widetilde{B}_{\mu\,
  m}$ from \eqref{11Dduals}.  This indicates why the $\omega$-rotation
may not be implementable in terms of the vielbein components
$e^m{}_{AB}$ and $e_{mn\,AB}$ only. We note that the above combination
breaks GL(7) invariance (and hence diffeomorphism invariance in the
internal dimensions); in fact, it just corresponds to the U(1)
rotation coming from the Ehlers $SL(2,{\mathbb{R}})$ symmetry which
enlarges the E$_{7(7)}$ of the $4D$ theory to the E$_{8(8)}$ symmetry
of $3D$ maximal supergravity.\\[.5ex]
%%%%%%%%%%%%%%%%%%
\noindent
Note added: Since this paper was submitted it has been shown that the
parameter $\lambda$ introduced in \eqref{eq:dual-vielbein-ansatz}
takes the universal value $\lambda=\tfrac12\sqrt{2}$ by considering
the non-linear flux formula in a variety of non-trivial backgrounds
\cite{Godazgar:2013nma}.

%%%%%%%%%%%%%%%%%%%%%%%%%%%%%%%%%%%%%%%%%%%%%%%%%%%%%%%%%%%%%%%%
\section*{Acknowledgements}
%%%%%%%%%%%%%%%%%%%%%%%%%%%%%%%%%%%%%%%%%%%%%%%%%%%%%%%%%%%%%%%%
We thank Gianluca Inverso, Mario Trigiante and Oscar Varela, as well as Hadi and Mahdi
Godazgar for valuable discussions. B.d.W. is supported by the ERC Advanced Grant
no. 246974, {\it ``Supersymmetry: a window to non-perturbative
  physics''}.
%%%%%%%%%%%%%%%%%%%%%%%%%%%%%%%%%%%%%%%%%%%%%%%%%%%%%%%%%%%%%%%%
%%%%%%%%%%%%%%%%%%%%%%%%%%%%%%%%%%%%%%%%%%%%%%%%%%%%%%%%%%%%%%%%

%%%%%%%%%%%%%%%%%%%%%%%%%%%%%%%%%%%%%%%%%%%%%%%%%%%%%%%%%%%%%%%
\begin{appendix}
%%%%%%%%%%%%%%%%%%%%%%%%%%%%%%%%%%%%%%%%%%%%%%%%%%%%%%%%%%%%%%%
\section{The supersymmetry variation of the dual generalized
    vielbein}
\label{App:derivation-dual-gen-vielbein}
\setcounter{equation}{0}
%%%%%%%%%%%%%%%%%%%%%%%%%%%%%%%%%%%%%%%%%%%%%%%%%%%%%%%%%%%%%%%
Here we present the evaluation of the supersymmetry transformation
\eqref{eq:susy-var-dual-vielbein} of the dual generalized vielbein,
defined in \eqref{eq:dual-gen-vielbein}. The derivation proceeds in
close analogy with the derivation of the third equation in
\eqref{eq:metric-KK} given originally in \cite{deWit:1986mz} (for
further details the reader is invited to consult eqs. (3.10) - (3.15) of that reference).

First let us summarize some of the definitions introduced in
\cite{deWit:1986mz}. The $11D$ fermion fields $\Psi_M$ and gamma
matrices $\tilde\Gamma_A$ are decomposed as, 
\begin{equation}
  \label{eq:11D-grav-decomp}
  \Psi_M(x,y) = \left\{\begin{array} {l} \Psi_\mu(x,y)\,, \\[.2ex]
      \Psi_m(x,y) \,, \end{array}\right.  \qquad
  \tilde\Gamma_A = \left\{\begin{array} {l} \tilde\Gamma_\alpha=
      \gamma_\alpha\otimes\oneone \,, \\[.2ex]
      \tilde\Gamma_a= \gamma_5\otimes \Gamma_a \,,
        \end{array}\right. 
\end{equation}
where the vielbeine $e_\mu{}^\alpha$ and $e_m{}^a$ have been defined
in \eqref{eq:elfbein} and the gamma matrices $\Gamma_a$ satisfy,
\begin{equation}
  \{ \Gamma_a,\,\Gamma_b\} = 2\,\delta_{ab} \,\oneone\,, \qquad
  \Gamma^{[a}\Gamma^b\cdots\Gamma^{g]} = -\mathrm{i}
  \varepsilon^{abcdefg}\,\oneone. 
\end{equation}
Furthermore the resulting chiral spinors, which carry upper and lower
$\mathrm{SU}(8)$ indices $A,B, \ldots$, are defined by
\begin{align}
  \label{eq:spinor-def}
  \psi_\mu{}^A(x,y)=&\, \tfrac12(1+\gamma_5)\,\mathrm{e}^{-\mathrm{i}\pi/4}\,
  \Delta^{1/4} \big(\Psi_\mu - B_\mu{}^m
  \Psi_m - \tfrac12\gamma_\mu \Delta^{-1/2}
  \Gamma^m\Psi_m\big)_A   \,,\nonumber\\
  \psi_{\mu A} (x,y)=&\, \tfrac12(1-\gamma_5)
  \,\mathrm{e}^{\mathrm{i}\pi/4}\, \Delta^{1/4} \big(\Psi_\mu - B_\mu{}^m
  \Psi_m + \tfrac12\gamma_\mu \Delta^{-1/2}
  \Gamma^m\Psi_m\big)_A  \,,\nonumber\\
  \epsilon^A(x,y) =&\,
  \tfrac12(1+\gamma_5)\,\mathrm{e}^{-\mathrm{i}\pi/4}\, \Delta^{1/4}
  \,\epsilon_A\,, 
  \nonumber\\
  \epsilon_A(x,y) =&\,
  \tfrac12(1-\gamma_5)\,\mathrm{e}^{\mathrm{i}\pi/4}\, \Delta^{1/4}
  \,\epsilon_A\,, 
  \nonumber\\
  \chi^{ABC} (x,y) =&\, \tfrac34\sqrt{2}\, (1+\gamma_5)
  \,\mathrm{e}^{-\mathrm{i}\pi/4}\, \Delta^{-1/4}  \mathrm{i}
  \Gamma_{a [AB} \Psi^a{}_{C]}  \,, \nonumber\\
\chi_{ABC} (x,y) =&\, \tfrac34\sqrt{2}\, (1-\gamma_5)
  \,\mathrm{e}^{\mathrm{i}\pi/4}\, \Delta^{-1/4}  \mathrm{i}
  \Gamma_{a [AB} \Psi^a{}_{C]}  \,,
\end{align}
where for the $11D$ spinors on the right-hand side we made no
distinction between upper and lower spinor indices and suppressed the
dependence on $x^\mu$ and $y^m$.

To derive the second equation in \eqref{eq:susy-var-dual-vielbein}, we
first evaluate the right-hand side of the equation, going `backwards'
from the $\mathrm{SU}(8)$ covariant expressions as in
\cite{Cremmer:1979up,deWit:1986mz}, but suppressing the
$\mathrm{SU}(8)$ compensating phase $\Phi$. Using SO(8) Fierz identities
given in \cite{deWit:1986mz}, we obtain in this way
\begin{align}
  \label{eq:compare-delta-e}
  -\sqrt{2}\Big(\bar\epsilon_{[A} \chi_{BCD} + & \tfrac1{24}
  \varepsilon_{ABCDEFGH}\, \bar\epsilon^{E}\chi^{FGH}\Big) \,
  e_{mn}{}^{CD} \nonumber\\
  = -\tfrac1{12}\mathrm{i}\sqrt{2} \,\Delta^{-1/2}\Big\{&-\tfrac1{32}
  \bar\epsilon(1-\gamma_5) (\Gamma_{bc}\Gamma_a \Gamma_{mn} +
  \Gamma_{mn}\Gamma_a\Gamma_{bc} )\Psi^a
  \;\Gamma^{bc}{}_{AB} \nonumber\\
  &+\tfrac1{16} \bar\epsilon(1-\gamma_5) (\Gamma_{b}\Gamma_a
  \Gamma_{mn} + \Gamma_{mn}\Gamma_a\Gamma_{b} )\Psi^a
  \;\Gamma^{b}{}_{AB} \nonumber\\
  &-\tfrac1{4} \bar\epsilon(1-\gamma_5) \Gamma_{mn}\Psi^a
  \;\Gamma_{a\, AB} \nonumber\\[.4ex]
  & -\tfrac1{12} \bar\epsilon(1+\gamma_5) \Gamma_{b[m} \Psi_{n]}
  \;\Gamma^b{}_{AB} \nonumber\\
  & -\tfrac1{12} \bar\epsilon(1+\gamma_5) \Gamma^{b} \Psi_{[m}
  \;\Gamma_{n]b\,AB} \nonumber\\
  & -\tfrac1{4} \bar\epsilon(1+\gamma_5) \Gamma_{[mn} \Psi^a
  \;\Gamma_{a]\,AB} \nonumber\\
  & -\tfrac1{4} \bar\epsilon(1+\gamma_5) \Gamma_{[a} \Psi^a
  \;\Gamma_{mn]\,AB} \nonumber\\
  & -\tfrac1{96} \bar\epsilon(1+\gamma_5) (\Gamma_{d[m} \Gamma^{bc}
  \Gamma^d +\Gamma^d \Gamma^{bc} \Gamma_{d[m})\Psi_{n]}
  \;\Gamma_{bc\,AB}
  \nonumber\\
  & +\tfrac1{48} \bar\epsilon(1+\gamma_5) (\Gamma_{d[m} \Gamma^{b}
  \Gamma^d +\Gamma^d \Gamma^{b} \Gamma_{d[m})\Psi_{n]}
  \;\Gamma_{b\,AB}
  \nonumber\\
  & -\tfrac1{32} \bar\epsilon(1+\gamma_5) (\Gamma_{[a} \Gamma^{bc}
  \Gamma_{mn]} +\Gamma_{[mn} \Gamma^{bc} \Gamma_{a]})\Psi^a
  \;\Gamma_{bc\,AB}
  \nonumber\\
  & +\tfrac1{16} \bar\epsilon(1+\gamma_5) (\Gamma_{[a} \Gamma^{b}
  \Gamma_{mn]} +\Gamma_{[mn} \Gamma^{b} \Gamma_{a]})\Psi^a
  \;\Gamma_{b\,AB} \Big\} \,,
\end{align}
This result should be compared to the left-hand side of the second
equation in \eqref{eq:susy-var-dual-vielbein}, which arises from the
variation of \eqref{eq:dual-gen-vielbein} as obtained from the $11D$
variations of the siebenbein $e_m{}^a$ and the three-form field
$A_{mnp}$,
\begin{align}
  \label{eq:original}
  & -\tfrac1{12}\mathrm{i}\sqrt{2} \big\{ \delta \big(\Delta^{-1/2} e_m{}^a
    e_n{}^b\big) \,\Gamma_{ab\, AB} 
  +6 \sqrt{2} \big(\delta A_{mnp}\big) \, \Delta^{-1/2}
    \,\Gamma^p{}_{AB}  \big\}  \nonumber\\
   & =-\tfrac1{12}\mathrm{i} \sqrt{2}\, \Delta^{-1/2}\big\{  -\bar\epsilon\gamma_5
   \Gamma^p\Psi_{[m} \;\Gamma_{n]p\,AB}  
  - \tfrac14 \bar\epsilon\gamma_5 \Gamma_a\Psi^a
  \;\Gamma_{mn\,AB} 
  -\tfrac32 \bar\epsilon \Gamma_{[mn} \Psi_{p]}
  \;\Gamma^p{}_{AB} \big\} \,.
\end{align}
where we suppressed the contribution proportional to $A_{mnp}
\,\delta\big( \Delta^{-1/2} e_a{}^p \big) \Gamma^a{}_{AB}$, as this
part of the variation is already taken care of by the calculation in
\cite{deWit:1986mz}, which corresponds to the result in the first line
of \eqref{eq:susy-var-dual-vielbein}.  As it turns out, the two
contributions \eqref{eq:compare-delta-e} and \eqref{eq:original} are
equal provided we add to \eqref{eq:original} an infinitesimal
$\mathrm{SU}(8)$ transformation acting on $e_{mn AB}$,
\begin{align}
  \label{eq:su8}
  \delta_{\mathrm{SU}(8)} e_{mn\, AB}= &\,  2\,\Lambda^C{}_{[A}\,
  e_{mn\, B]C}\nonumber\\
  =&\, -\tfrac1{12}\mathrm{i}\sqrt{2}\, \Delta^{-1/2} \Big[
  -\tfrac18 \bar\epsilon \Gamma^{ab}\Psi^{c}\; \Gamma_{abcmn\,AB}
  +\tfrac34 \bar\epsilon\Gamma_{[mn}\Psi_{a]} \;\Gamma^{a}_{AB} \\
  & \qquad\qquad -\tfrac12\bar\epsilon\gamma_5 \Gamma_{a[m}\Psi^a\; \Gamma_{n]\,AB}
  +\tfrac12 \bar\epsilon\gamma_5 \Gamma^a \Psi_{[m} \;\Gamma_{n]a\,AB}
  - \tfrac12 \bar\epsilon\gamma_5 \Gamma_{[m} \Psi^a
  \;\Gamma_{n]a\,AB}\Big] \,,\nonumber
\end{align}
where the parameter of this transformation takes the form 
\begin{equation}
\label{Lambda}
\Lambda_A{}^B = - \Lambda^B{}_A 
\equiv \tfrac18 \bar\epsilon \gamma_5 \Gamma_{ab} \Psi^b \Gamma^a_{AB}
  - \tfrac18 \bar\epsilon \gamma_5 \Gamma_a \Psi_b \Gamma^{ab}_{AB}
  - \tfrac1{16} \bar\epsilon \Gamma_{ab} \Psi_c \Gamma^{abc}_{AB}\,. 
\end{equation}
The expression for the $\mathrm{SU}(8)$ parameter \eqref{Lambda} is
identical to the one given in eq.~(3.13) of \cite{deWit:1986mz}, where
it was found by determining the $\mathrm{SU}(8)$ covariant form of the
supersymmetry transformation of $e^m{}_{AB}$. This remarkable
coincidence is not only crucial for the correctness of the second
equation \eqref{eq:susy-var-dual-vielbein}, but it is also another
non-trivial consistency check of the $\mathrm{SU}(8)$ invariant
reformulation of $11D$ supergravity presented in \cite{deWit:1986mz}.

%%%%%%%%%%%%%%%%%%%%%%%%%%%%%%%%%%%%%%%%%%%%%%%%%%%%%%%%%%%%%%%
\end{appendix}

%%%%%%%%%%%%%%%%%%%%%%%%%%%%%%%%
%\begin{thebibliography}{99}
\providecommand{\href}[2]{#2}
%\begingroup\raggedright

%%%%%%%%%%%%%%%%%%%%%%%%%%%%%%%%%%%%%%%%%%%%%%%%%%%%%%%%%%%%%%%%%%
\end{document}